\documentclass[a4paper,10pt,notitlepage]{article}
\usepackage{amsfonts,amssymb,amsmath,amsthm,hyperref,colonequals}
\usepackage[mathscr]{euscript}
\usepackage{bbm} % Used for the identity symbol
\usepackage[show]{ed} % Used for putting comments in the margins
\usepackage{makeidx}
\usepackage{graphicx}
\usepackage{url}
\usepackage{multirow}
\usepackage{youngtab}
\usepackage[arrow, matrix, curve]{xy}
\theoremstyle{definition}
\newcommand{\beqa}{\begin{eqnarray}}
\newcommand{\eeqa}{\end{eqnarray}}
\newcommand{\beq}{\begin{equation}}
\newcommand{\eeq}{\end{equation}}

\DeclareMathOperator{\g}{\mathfrak{g}}
\newcommand{\osp}[2]{\ensuremath{\mathfrak{osp}\left({#1}|{#2}\right)}}
\newcommand{\psl}[1]{\ensuremath{\mathfrak{psl}\left({#1}|{#1}\right)}}

\DeclareMathOperator{\G}{\mathrm{G}}

\newcommand{\BU}[1]{\ensuremath{\text{U}\left(\text{{#1}}\right)}}

\newcommand{\calF}{\mathcal{F}}\newcommand{\calH}{\mathcal{H}}
\newcommand{\calS}{\mathcal{S}}

\newcommand{\calB}{\mathcal{B}}

\newcommand{\calO}{\mathcal{O}}\newcommand{\calJ}{\mathcal{J}}

\newcommand{\calD}{\mathcal{D}}

\newcommand{\fg}{\mathfrak{g}}
\newcommand{\tg}{\textsf{g}}

\newcommand{\bT}{\mathbf{T}}\newcommand{\bK}{\mathbf{K}}

\newcommand\str{\text{str}}
\newcommand\sdim{\text{sdim}\,}
\DeclareMathOperator{\Cas}{\mathbf{Cas}}

\newcommand{\vac}[1]{\ensuremath{\left< \, #1\, \right>}}

\newcommand\dil{\Delta}
\newcommand\met{\mathscr{I}}

\begin{document}
%%%%%%%%%%%%%%%%%%%%%%%%%%%%%%%%%%%%%%%%%%%%%%%%%%%%%%%%%%%%%%%%%%%%

%%%%%%%%%%%%%%%%%%%%%% TITLE PAGE %%%%%%%%%%%%%%%%%%%%%%%%%%%%%%%%%%%
\thispagestyle{empty}
\setcounter{page}{0}
\begin{flushright}\footnotesize
\texttt{DESY 12-186}\\
\texttt{HU-Mathematik-11-2012}\\
\texttt{HU-EP-12/34}\\
\vspace{0.5cm}
\end{flushright}
\setcounter{footnote}{0}

\begin{center}
{\Large\textbf{\mathversion{bold}
Anomalous Dimensions in \\[2mm] Deformed WZW Models on Supergroups
}\par}
\vspace{15mm}

{\sc  Constantin Candu $^{a}$,
Vladimir Mitev $^{b}$,   Volker Schomerus $^{c}$ ,}\\[5mm]

{\it $^a$ Institut f\"ur Theoretische Physik,  HIT K 21.5, Wolfgang-Pauli-Str. 27\\
8093 Z\"urich, Switzerland
}\\[5mm]

{\it $^b$ Institut f\"ur Mathematik und Institut f\"ur Physik,\\ Humboldt-Universit\"at zu Berlin\\
IRIS Haus, Zum Gro{\ss}en Windkanal 6,  12489 Berlin, Germany
}\\[5mm]

{\it $^c$ DESY Hamburg, Theory Group, \\
Notkestrasse 85, D--22607 Hamburg, Germany
}\\[5mm]

\texttt{canduc@itp.phys.ethz.ch}\\
\texttt{mitev@math.hu-berlin.de}\\
\texttt{volker.schomerus@desy.de}\\[25mm]

\textbf{Abstract}\\[2mm]
\end{center}
We investigate a class of current-current, Gross-Neveu like, perturbations of WZW models in which the full left-right affine symmetry is broken to the diagonal global algebra only. Our analysis focuses on those supergroups for which such a perturbation preserves conformal invariance. A detailed calculation of the 2-point functions of affine primary operators to 3-loops is presented. Furthermore, we derive an exact formula for the anomalous dimensions of a large subset of fields to all orders in perturbation theory. Possible applications of our results, including the study of non-perturbative
dualities, are outlined.

%%%%%%%%%%%%%%%%%%%%%%%%% END OF TITLE PAGE %%%%%%%%%%%%%%%%%%%%%%%%%%%%%

\newpage
\setcounter{page}{1}

%%%%%%%%%%%%%%%%%%%%%%%%%%%%%%%%%%%%%%%%%%%%%%%%%%%%%%%%%%%%%%%%%%%%

\tableofcontents
\addtolength{\baselineskip}{5pt}

%%%%%%%%%%%%%%%%%%%%%%%%%%%%%%%%%%%%%%%%%%%%%%%%%%%%%%%%%%%%%%%%%%%%
%%%%%%%%%%%%%%%%%%%%%%%%%%%%%%%%%%%%%%%%%%%%%%%%%%%%%%%%%%%%%%%%%%%%

\section{Introduction}\label{sec:intro}

%%%%%%%%%%%%%%%%%%%%%%%%%%%%%%%%%%%%%%%%%%%%%%%%%%%%%%%%%%%%%%%%%%%%
%%%%%%%%%%%%%%%%%%%%%%%%%%%%%%%%%%%%%%%%%%%%%%%%%%%%%%%%%%%%%%%%%%%%

Conformal field theories (CFT) with internal supersymmetry are quite
relevant for a number of different problems in modern mathematical
physics, from strongly correlated electronic systems \cite{Efetov:1983xg, Bernard:1995as, Zirnbauer:1999ua}, to string theory \cite{Berkovits:1999im, Bershadsky:1999hk}. One class of examples is obtained from Wess-Zumino-Witten
(WZW) models on Ricci-flat (simple) supergroups $\G$.
These possess a marginal (current-current) deformation which preserves the (diagonal) global $\G$ symmetry and conformal invariance, see \cite{Mitev:2008yt,Konechny:2010nq} and references in \cite{Candu:2011hu}. There are a number of important incarnations of such theories, e.g.\ as supergroup Gross-Neveu models or -- closely related -- as a fermionic sector in supergroup sigma models with world-sheet supersymmetry. Sigma models on certain symmetric
superspaces $\G/\,\mathrm{H}$, which have been classified in~\cite{Candu:thesis, Candu:2010yg}, provide another class of examples for CFTs with target space supersymmetry. In the present work, we shall concentrate on current-current deformations of WZW models. Sigma models will be the focus of a forthcoming publication.

Previous studies of CFTs with internal supersymmetry have uncovered a number of remarkable features. In particular, it was argued in \cite{Bershadsky:1999hk,Quella:2007sg} that the perturbative evaluation of certain special quantities, such as conformal weights, may be insensitive to the non-abelian nature of the symmetry $\G$, provided $\G$ is Ricci-flat. In those cases it may then be possible to sum up the entire perturbative series. The best studied examples of such phenomena appear when the WZW model is placed on the upper half plane with symmetry preserving boundary conditions. In the
case of the OSP(4$|2$) WZW model at level $k=1$, for example, a proposal for the exact boundary partition function was put forward in \cite{Mitev:2008yt}. The anomalous dimensions of boundary fields in this proposal are given by the value of the quadratic Casimir of the representation in which the fields transform, multiplied by a \textit{universal} field independent function. This simple behavior is compatible with numerical studies of an appropriate
$\osp{4}{2}$ spin chain, see \cite{Candu:2008vw,Candu:2008yw}. The exact formula for the boundary partition function made possible to confirm \cite{Mitev:2008yt} a conjectured non-perturbative duality between the $\osp{4}{2}$ Gross-Neveu model and the sigma models on the supersphere $S^{3|2}$, see \cite{Candu:2008yw}.

Given the powerful applications that exact results for the anomalous dimensions may have, it seems worthwhile to extend their perturbative computation to the bulk theory. This is the main goal of our present work. Note that bulk spectra are usually much more complex than those associated with excitations of a boundary. In the compactified free boson, for example, the bulk spectrum contains both momentum and winding number while on the boundary only one of them appears, depending on whether we impose Dirichlet or Neumann boundary
conditions. The complexity of bulk spectra may seem like an obstacle at first, but as we shall show below,  it actually turns into a virtue.

Before we describe the main results of this work and outline the plan, let us recall that supergroup WZW models are almost always logarithmic \cite{Rozansky:1992rx,Schomerus:2005bf,Quella:2007hr}, meaning that the dilation operator is not diagonalizable. In order to spell out the general structure of 2-point functions in such models, let us
consider two field multiplets $\Phi = (\phi_a)$ and $\Psi = (\psi_b)$, which we can formally view as forms on the carrier spaces $V_\Phi$ and $V_\Psi$ of two
indecomposable (but not necessarily irreducible) representations of the
Lie superalgebra $\g$. Conformal symmetry implies that
\begin{equation} \label{twopt}
\langle \Phi(u) \Psi(v) \rangle =  \mathscr{I} \circ |u-v|^{-\dil_\Phi - \dil_\Psi}\ .
\end{equation}
Here, $\dil_\Phi + \dil_\Psi \equiv \dil_\Phi \otimes \text{id}_{\Psi} + \text{id}_{\Phi} \otimes
\dil_\Psi$ is built from the representations $\dil_\Phi$ and $\dil_\Psi$ of the dilation operator on the field multiplets $\Phi$ and $\Psi$, respectively and the symbol $\met$ denotes an intertwiner $\met:V_\Phi \otimes V_\Psi \rightarrow \mathbb{C}$ from the tensor product of the two multiplets into the scalars.\\
If $V_\Phi$ and $V_\Psi$ are irreducible, $\dil_\Phi$ and
$\dil_\Psi$ are diagonal and the intertwiner $\met$ is unique up to an overall factor that can be absorbed in a normalization of fields. Hence, we recover the usual form of a 2-point function without logarithms. In perturbation theory, the intertwiner $\met$ can acquire a non-trivial scalar factor which may be reabsorbed in the normalization of the fields. Hence, the main interest is to compute how the eigenvalue of $\dil_\psi$ changes with the deformation. \\
Atypical representations (short multiplets) can form complicated
indecomposables which are not irreducible. In such cases, there often exist several independent intertwiners $\met$ and, moreover, the representation of the dilation operator $\dil_\Phi$ may no longer be diagonalizable. Whenever this happens, the formula \eqref{twopt} contains terms with a logarithmic dependence on the difference of the world-sheet coordinates $u$ and $v$. The coefficients of these logarithmic terms are physical, i.e.\ they can not be absorbed by renormalizing the fields.

Our main goal in this article is to analyze the 2-point functions, and in
particular the behavior of anomalous dimensions, of bulk fields in
deformed WZW models. In a first step we shall compute the exact bulk
(and boundary) 2-point functions for all affine primaries up to 3-loops. Through explicit evaluation of the appearing
integrals we shall show that, to 3-loop order, the only effect of a
current-current perturbation is to change the conformal weight of affine primaries. Moreover, the anomalous contribution to the conformal weight is determined through a universal formula from the eigenvalues of quadratic Casimir elements. The precise expression for bulk fields can be found in eq.~\eqref{eq:3loopanomalousdimensions} and for boundary fields in eq.~\eqref{eq:anomalousdimboundary}.
As we reviewed above, the 2-point function of fields in WZW models
on supergroups may contain logarithms.
These logarithmic terms are also computed up to 3-loops. In particular, if logarithms are absent from the 2-point function of affine primaries in the unperturbed model, they will not appear before the fourth order in perturbation theory. It might be possible to push
this result to even higher loop order, but this will require some tedious work.

Having discussed the exact 2-point functions for bulk- and boundary
fields in section \ref{sec:twopointfun3loops}, we shall then turn to the maximally atypical -- or $\tfrac{1}{2}$BPS in physics terminology -- sector with respect to the the diagonal symmetry of the bulk model. By definition, all irreducible multiplets in this subsector possess non-zero superdimension. For such $\tfrac{1}{2}$BPS fields we shall show that the perturbation theory becomes quasi-abelian. More precisely, all terms in the perturbation series that contain the structure constants of the superalgebra vanish due to simple group theoretic identities, see section \ref{subsec:restrictionsonfields}. The resulting quasi-abelian perturbation series is summed up in section \ref{subsec:abelpertbulk}. We find that the conformal weights of fields within the maximally atypical sector evolve with the quadratic Casimir of the left, right and diagonal group actions in the unperturbed WZW model. The precise spectrum of $\tfrac{1}{2}$BPS states depends on the details of the model, i.e. on the modular invariant partition function we start with before turning on the perturbation. Given any particular
model, we can determine the $\tfrac{1}{2}$BPS spectrum and then, using the
results from this paper, evolve this spectrum to any point on the
1-parameter parameter (moduli) space of the current-current
deformation. Since the $\tfrac{1}{2}$BPS sector can be considered as a footprint of the model, our results should provide a valuable new tool for discovering dualities in the space of CFTs with target space supersymmetry. We shall discuss the issue further in the concluding section of this article.

%%%%%%%%%%%%%%%%%%%%%%%%%%%%%%%%%%%%%%%%%%%%%%%%%%%%%%%%%%%%%%%%%%%%
%%%%%%%%%%%%%%%%%%%%%%%%%%%%%%%%%%%%%%%%%%%%%%%%%%%%%%%%%%%%%%%%%%%%

\section{The deformed WZW model}

%%%%%%%%%%%%%%%%%%%%%%%%%%%%%%%%%%%%%%%%%%%%%%%%%%%%%%%%%%%%%%%%%%%%
%%%%%%%%%%%%%%%%%%%%%%%%%%%%%%%%%%%%%%%%%%%%%%%%%%%%%%%%%%%%%%%%%%%%

Consider a WZW model on a simple supergroup $\G$ with current algebra symmetries
\begin{equation}\label{eq:currentsOPE}
J^a(z)J^b(w)\sim \frac{k \eta^{ab}}{(z-w)^2} + \frac{{f^{ab}}_c J^c(w)}{z-w},\quad
\bar{J}^a(\bar{z})\bar{J}^b(\bar{w})\sim \frac{k \eta^{ab}}{(\bar{z}-\bar{w})^2} + \frac{{f^{ab}}_c \bar{J}^c(\bar{w})}{\bar{z}-\bar{w}}\ .
\end{equation}
Here  ${f^{ab}}_c$ are the structure constants of the Lie algebra $\fg$ of $\G$, $\eta^{ab}$ is  an invariant bilinear form
and  $k$ is the level of the current algebra (normalized with respect $\eta^{ab}$).
The full symmetry of the model is two copies $\fg_k\times \fg_k$ of the affine algebra $\fg$ at level $k$ represented by the modes of the holomorphic
and antiholomorphic currents
\begin{equation}%\label{eq:}
J^a(z)=\sum_{n\in\mathbb{Z}} J^a_n z^{-n-1}, \qquad \bar{J}^a(\bar{z})=\sum_{n\in\mathbb{Z}} \bar{J}^a_n \bar{z}^{-n-1}.
\end{equation}
We shall make no assumption about the bulk spectrum of the WZW model.

Next, let us perturb the WZW model over the Riemann sphere by a marginal current-current term breaking the full symmetry group $\G_k\times \G_k$ of the original model down to the global diagonal subgroup. Its action reads
\begin{equation}
\label{eq:pwzw}
\calS \colonequals \calS_0 +\calS_{\text{int}}=\calS_0+ g \int \frac{d^2 z}{\pi}\, J^a(z)\bar{J}^b(\bar{z}) \eta_{ab}\ ,
\end{equation}
where $\calS_0$ denotes the action of the WZW model and $\eta_{ab}$ is the inverse of $\eta^{ab}$.
It was proven in~\cite{Bershadsky:1999hk, Berkovits:1999im, Candu:thesis, Candu:2010yg} that the perturbed theory~\eqref{eq:pwzw} remains conformally invariant whenever the (complexified) Lie
algebra of $\G$ is taken from the list
\begin{equation}\label{eq:gtype}
\psl{n}\ ,\quad \osp{2n+2}{2n}\ ,\quad  D(2,1;\alpha)\ .
\end{equation}
Our goal  is to compute the spectrum of anomalous dimensions in the bulk for the
conformal perturbations~\eqref{eq:pwzw} of WZW models with $\fg$ of the type~\eqref{eq:gtype}.
We shall extract them from the analysis of the 2-point function
of some arbitrary fields $\Phi$, $\Psi$, which
can be computed perturbatively  in the usual way

\begin{equation}\label{eq:2pf}
G(u,v)\colonequals
\vac{\Phi(u,\bar{u})\Psi(v,\bar{v})}=\vac{\Phi(u,\bar{u})\Psi(v,\bar{v})e^{-\calS_{\text{int}}}}_{0,c}=
\sum_{n=0}^\infty G^{(n)}(u,v)\ .
\end{equation}
Here $\langle\cdot \rangle_0$ denotes a correlation function in the original theory, $c$ stands for connected diagrams with vacuum bubbles removed
and $G^{(n)}(u,v)$ is the $n$-loop contribution
\begin{equation}\label{eq:nG}
G^{(n)}(u,v)
\colonequals \frac{(-g)^n}{n!}
\int_{\mathcal{D}_n}\frac{d^2z_1\cdots d^2 z_n}{\pi^n}\, \langle \Phi(u,\bar{u})\Psi(v,\bar{v})\Omega(z_1,\bar{z}_1)\cdots \Omega(z_n,\bar{z}_n)\rangle_0 \ ,
\end{equation}
where  $\Omega(z,\bar{z})\colonequals  J^a(z)\bar{J}^b(\bar{z}) \eta_{ab}$.
The integrals will be regularized by a short distance cut-off~$\epsilon$, so that the integration domain becomes
\begin{equation}
\label{eq:defdomainD}
\calD_n \colonequals \big\{(z_1,\ldots,z_n)\in \mathbb{C}^n\ :\ |z_i-u|>\epsilon\ ,\; |z_i-v|>\epsilon\ ,\; |z_i-z_j|>\epsilon\big\}\ .
\end{equation}
The correlation functions of the unperturbed theory in eq.~\eqref{eq:nG} can be computed with the help of the Ward identities~\footnote{ For simplicity, we assume that $O_i$ are bosonic so that there are no grading signs.} for the holomorphic currents $J^a$
\begin{equation}\label{eq:ward}
\vac{J^a(z) O_1(z_1)\cdots O_n(z_n)}_0 =\sum_{i=1}^n
\vac{ O_1(z_1)\cdots [J^a(z)O_i(z_i)]\cdots O_n(z_n)}_0
\end{equation}
and their antiholomorphic counterparts $\bar{J}^a$, where $[J^a(z)O_i(z_i)]$ denotes the singular part of the corresponding OPE.
Thus, if the OPEs of the operators $O_i(z_i)$ with the currents are known, then the above equation tells us how to reduce their $(n+1)$-point
correlation functions with a current insertion
to (sums of) $n$-point correlation functions.
One can apply these identities recursively in order to express the $(n+2)$-point correlation function in eq.~\eqref{eq:nG} in terms of 2-point functions only.

%%%%%%%%%%%%%%%%%%%%%%%%%%%%%%%%%%%%%%%%%%%%%%%%%%%%%%%%%%%%%%%%%%%%
%%%%%%%%%%%%%%%%%%%%%%%%%%%%%%%%%%%%%%%%%%%%%%%%%%%%%%%%%%%%%%%%%%%%

\section{The structure of the 2-point function}
\label{sec:pert}
%%%%%%%%%%%%%%%%%%%%%%%%%%%%%%%%%%%%%%%%%%%%%%%%%%%%%%%%%%%%%%%%%%%%
%%%%%%%%%%%%%%%%%%%%%%%%%%%%%%%%%%%%%%%%%%%%%%%%%%%%%%%%%%%%%%%%%%%%

The CFTs we are dealing with are \emph{logarithmic}.
We shall recall some of their features which are necessary
to understand the basic structure of their 2-point function.
A field $\Phi(z,\bar{z})$ is called (quasi)primary if the
(global) conformal symmetry generators  act on it as
\begin{equation}%\label{eq:}
[L_n,\Phi(z,\bar{z})] = (z^n\boldsymbol{h}+z^{n+1}\partial )\cdot \Phi(z,\bar{z})\ ,\quad  [\bar{L}_n,\Phi(z,\bar{z})] = (\bar{z}^n\boldsymbol{h}+\bar{z}^{n+1}\bar{\partial} )\cdot \Phi(z,\bar{z})\ ,
\end{equation}
where $(\boldsymbol{h},\bar{\boldsymbol{h}})$ are the \emph{operator} conformal dimensions of $\Phi$.
If $\Phi$ is part of an indecomposable representation of $\g$, then $(\boldsymbol{h},\bar{\boldsymbol{h}})$ are  matrices, not necessarily diagonalizable, acting on  the fields of this representation.
As they commute with the action of $\g$, Schur's lemma restricts their form to
\begin{equation}%\label{eq:}
\boldsymbol{h} = h \boldsymbol{1} + \boldsymbol{h}_n\ ,
\end{equation}
where $h$ is the  conformal dimension (in the usual sense), while $\boldsymbol{h}_n, \bar{\boldsymbol{h}}_n$ are nilpotent matrices.~\footnote{ The nilpotent pieces can be non-zero only if the representation is not irreducible.}
Locality requires $\boldsymbol{h}_n=\bar{\boldsymbol{h}}_n$.
Moreover, by global conformal symmetry,  the 2-point function of two quasi-primaries $\Phi_1$ and $\Phi_2$ can be non-zero only if $\boldsymbol{h}_1-\boldsymbol{h}_2, \bar{\boldsymbol{h}}_1-\bar{\boldsymbol{h}}_2\in \mathbb{Z} \boldsymbol{1}$, which in particular means that the nilpotent pieces must agree.

Returning to the 2-point function of eq.~\eqref{eq:2pf},
let us  now explain the standard algorithm that one must follow in order
to extract the anomalous dimensions from the perturbative expansion.
Suppose we start with some fields $\Phi^{(0)}$ and $\Psi^{(0)}$,
which are quasi-primary in the WZW model and both of operator conformal dimension $(\boldsymbol{h}^{(0)},\bar{\boldsymbol{h}}^{(0)})$.
Then, already at 1-loop, the perturbative corrections to $\langle \Phi^{(0)}(u,\bar{u}) \Psi^{(0)}(v,\bar{v}) \rangle$ will typically contain divergences.
These can have either a power like, logarithmic, or mixed functional behavior of the cut-off $\epsilon$.
The power like and mixed divergences must be removed by \emph{cut-off dependent} field redefinitions
\begin{equation}\label{eq:fred}
\Phi = \sum_{n=0}^\infty \Phi^{(n)}\ ,\qquad \Psi = \sum_{n=0}^\infty \Psi^{(n)}\ ,
\end{equation}
where the $n$-th term in the sum denotes the $n$-loop $\mathcal{O}(g^n)$ correction.
The remaining logarithmically divergent and regular terms are then interpreted as matrix elements of the matrix of anomalous dimensions.
The eigenvectors of this matrix are fields with definite conformal dimensions in the interacting theory, while the  eigenvalues give their anomalous dimensions.~\footnote{
Here eigenvectors and eigenvalues are meant in a generalized sense:
$v$ is called a generalized eigenvector of a matrix $M$ with generalized eigenvalue $\lambda$ if there is a positive integer $n$ such that $(M-\lambda)^n\cdot v=0$.
}

Suppose now that the fields $\Phi, \Psi$ given in the expression~\eqref{eq:fred} are quasi-primary in the interacting theory, both of operator conformal dimensions $(\boldsymbol{h},\bar{\boldsymbol{h}})$.
Global conformal and $\g$-invariance restricts their 2-point function to be of the form
\begin{equation}\label{eq:2pfint}
G(u,v)=\vac{\Phi(u,\bar{u})\Psi(v,\bar{v})} = d\left(\Phi,\frac{\epsilon^{4{\boldsymbol\delta}}}{(u-v)^{2{\boldsymbol{h}}}(\bar{u}-\bar{v})^{2\bar{\boldsymbol{h}}}}\Psi\right)\ ,
\end{equation}
where $d(\Phi,\Psi)$ is a $\g$-invariant metric on the space of fields of the interacting theory and
\begin{equation}%\label{eq:}
\boldsymbol{\delta}\colonequals \boldsymbol{h} - \boldsymbol{h}^{(0)}= \bar{\boldsymbol{h}} - \bar{\boldsymbol{h}}^{(0)} =\delta\boldsymbol{1}+ \boldsymbol{\delta}_n
\end{equation}
is the operator anomalous dimension of the (local) fields $\Phi$ and $\Psi$.
We have decomposed it into a diagonal and a nilpotent piece, with $\delta$ being the usual anomalous dimension.

The metric $d$ appearing in eq.~\eqref{eq:2pfint} will in general differ from that of the original theory
\begin{equation}
\label{eq:defG0}
G^{(0)}(u,v) =\langle\,\Phi(u,\bar{u})\Psi(v,\bar{v})\,\rangle_0 =d^{(0)}\left(\Phi,\frac{1}{(u-v)^{2{\boldsymbol{h}^{(0)}}}(\bar{u}-\bar{v})^{2\bar{\boldsymbol{h}}^{(0)}}}\Psi\right)\ .
\end{equation}
Assuming that $\Psi$ is part of an indecomposable representation of $\g$, while $\Phi$ of its dual, one can write the most general form of $d$ in terms of $d^{(0)}$ as
\begin{equation}%\label{eq:}
d (\Phi,\Psi) = d^{(0)}(\Phi, \boldsymbol{a}\Psi)\ ,
\end{equation}
where $\boldsymbol{a}$ is an invertible matrix commuting with the action of $\g$.
Again, by Schur lemma, the latter must decompose into a diagonal and a nilpotent piece
\begin{equation}%\label{eq:}
\boldsymbol{a} = a\cdot \boldsymbol{1} + \boldsymbol{a}_n\ .
\end{equation}

The correlation function~\eqref{eq:2pfint} will appear in perturbation theory as a
power series of logarithmically divergent terms of the form
\begin{equation}\label{eq:2ptoan}
G(u,v) = \boldsymbol{a}\left(\boldsymbol{1} + 2 \boldsymbol{\delta} \log\frac{\epsilon^2}{|u-v|^2}+\cdots \right)\cdot G^{(0)}(u,v)\ ,
\end{equation}
where the dots denote higher powers of the logarithm.
Eq.~\eqref{eq:2ptoan} gives a prescription for  determining the form of quasi-primary fields
of the interacting theory. More precisely, it determines the field redefinitions~\eqref{eq:fred}
of the WZW model quasi-primaries  which bring the 2-point function $\langle \Phi^{(0)}(u,\bar{u})\Psi^{(0)}(v,\bar{v})\rangle$ to the canonical form of eq.~\eqref{eq:2ptoan}.

%%%%%%%%%%%%%%%%%%%%%%%%%%%%%%%%%%%%%%%%%%%%%%%%%%%%%%%%%%%%%%%%%%%%
%%%%%%%%%%%%%%%%%%%%%%%%%%%%%%%%%%%%%%%%%%%%%%%%%%%%%%%%%%%%%%%%%%%%
\section{The 2-point function at 3-loops}
\label{sec:twopointfun3loops}
%%%%%%%%%%%%%%%%%%%%%%%%%%%%%%%%%%%%%%%%%%%%%%%%%%%%%%%%%%%%%%%%%%%%%%%%
%%%%%%%%%%%%%%%%%%%%%%%%%%%%%%%%%%%%%%%%%%%%%%%%%%%%%%%%%%%%%%%%%%%%

The computation of  the 2-point functions simplifies considerably if we restrict to affine primaries. This is due to the fact that, at least for a generic level $k$, their conformal dimensions do not allow them to mix with the other fields, or, more precisely, because the 2-point function of an affine primary with any other field which is not an affine primary in the same affine conformal tower must vanish. Thus,  affine primaries should become primaries of the interacting theory without any field redefinition of the kind presented in eq.~\eqref{eq:fred} being required. For the same reason, the ``matrix of anomalous dimensions'' must be an operator built out of the zero modes of the currents, at least when restricted to the space of affine primaries.

In this section, we shall present in detail, order by order, the computation of the 2-point functions of affine primaries up to 3-loops, in the bulk as well as in the boundary theories. Furthermore, we also compute the anomalous dimensions of all quasi-primary fields to 1-loop. For the bulk affine primary fields, we find that at 3-loops there is a contribution to the integrand coming from the structure constants, but the latter turns out to vanish after integration. We also show that in the boundary theory structure constants start contributing to the integrand from 4-loops on.

%%%%%%%%%%%%%%%%%%%%%%%%%%%%%%%%%%%%%%%%%%%%%%%%%%%%%%%%%%%%%%%%%%%%
\subsection{Bulk affine primaries}
%%%%%%%%%%%%%%%%%%%%%%%%%%%%%%%%%%%%%%%%%%%%%%%%%%%%%%%%%%%%%%%%%%%%%%%%

The defining OPEs of a local affine primary operator $\Phi(w,\bar{w})$ with the currents are
\begin{equation}\label{eq:affpr}
J^a(z)\Phi(w,\bar{w})\sim \frac{(J^a_0\Phi)(w,\bar{w})}{z-w}\ ,\qquad \bar{J}^a(\bar{z})\Phi(w,\bar{w})\sim \frac{(\bar{J}^a_0\Phi)(w,\bar{w})}{\bar{z}-\bar{w}}\ .
\end{equation}
In the following we shall pull out of correlation functions the zero modes as follows
\begin{align}%\label{eq:}
L^a_\Phi\cdot \langle \Phi(z,\bar{z})\cdots \rangle_0 & =
 \langle (J^a_0\Phi)(z,\bar{z})\cdots \rangle_0\ ,&
R^a_\Phi\cdot \langle \Phi(z,\bar{z})\cdots \rangle_0 &=
\langle (\bar{J}^a_0\Phi)(z,\bar{z})\cdots \rangle_0\ .
\end{align}
Now, let $\Phi, \Psi$ be ground states of the same affine conformal tower.
We shall compute their 2-point function $G(u,v)$, introduced in eq.~\eqref{eq:2pf}, order by order. Using the Ward identities in eq.~\eqref{eq:ward} and the OPEs of eq.~\eqref{eq:affpr}, we can express the integrand of the 1-loop correction spelled out in eq.~\eqref{eq:nG} as
\begin{equation}\label{eq:1lintegrand}
\langle \Omega(z_1,\bar{z}_1)\Phi(u,\bar{u})\Psi(v,\bar{v})\rangle_{0,c}=\eta_{ab}\left(\frac{L^a_{\Phi}}{z_1-u}+\frac{L^a_{\Psi}}{z_1-v}\right)\left(\frac{R^b_{\Phi}}{\bar{z}_1-\bar{u}}+\frac{R^b_{\Psi}}{\bar{z}_1-\bar{v}}\right)\cdot G^{(0)}(u,v)\ ,
\end{equation}
where $G^{(0)}(u,v)\colonequals \langle \Phi(u,\bar{u}) \Psi(v,\bar{v})\rangle_0$.
Next, taking into account the global $\fg$ invariance of the correlation function
\begin{equation}\label{eq:inv1}
(L^a_{\Phi}+L^a_\Psi)\cdot G^{(0)}(u,v)=(R^a_\Phi+ R^a_\Psi)\cdot G^{(0)}(u,v)=0 \
\end{equation}
we get for the 1-loop contribution an expression of the form
\begin{equation}\label{eq:1laffcompact}
G^{(1)}(u,v)= -g  \mathcal{I}(u,v)\eta_{ab} L^a_\Phi R^a_\Phi\cdot G^{(0)}(u,v)\ ,
\end{equation}
containing a 1-loop integral
\begin{equation}\label{eq:int1}
\mathcal{I}(u,v):=\int_{\mathcal{D}_1} \frac{d^2z_1}{\pi}\, \left( \frac{1}{z_1-u}-\frac{1}{z_1-v}\right) \left( \frac{1}{\bar{z}_1-\bar{u}} -\frac{1}{\bar{z}_1-\bar{v}}\right)=
\int_{\mathcal{D}_1} \frac{d^2z_1}{\pi}\, h_1\bar{h}_1
\ .
\end{equation}
In the above, we have introduced a shorthand for the function
\begin{equation}
\label{eq:deffunctionh}
h_i\equiv h_i(u,v)\colonequals \frac{1}{z_i-u}-\frac{1}{z_i-v}\ .
\end{equation}
that we shall encounter frequently.
The integral of eq.~\eqref{eq:int1} is computed in app.~\ref{sec:int}, with the result
\begin{equation}%\label{eq:}
\mathcal{I}(u,v)=2\log \frac{|u-v|^2}{\epsilon^2}+\mathcal{O}(\epsilon^2)\ .
\end{equation}
The term in eq.~\eqref{eq:1laffcompact} that is bilinear in the zero modes of the currents can be assembled into Casimirs as follows.
Evaluating the Casimir  $\Cas\colonequals T^aT^b\eta_{ab}$ of the superalgebra $\fg$ in the left, right and diagonal representations carried by $\Phi$ we get three Casimir operators
\begin{equation}%\label{eq:}
\Cas_{L}\colonequals L_{\Phi}^a L_{\Phi}^b\eta_{ab}, \quad \Cas_{R}\colonequals R_{\Phi}^aR_{\Phi}^b\eta_{ab}, \quad  \Cas_{D}\colonequals(L_{\Phi}^a+R_{\Phi}^a)(L_{\Phi}^b+R_{\Phi}^b)\eta_{ab}\ .
\end{equation}
Therefore, the operator appearing in eq.~\eqref{eq:1laffcompact} can be rewritten as
\begin{equation}%\label{eq:}
\label{eq:definitionofT}
\bT\colonequals \eta_{ab}L_{\Phi}^a R_{\Phi}^b=\tfrac{1}{2}\left(\Cas_{D}-\Cas_{L}-\Cas_{R}\right)\ .
\end{equation}
Putting these results together, we arrive at a 2-point function of the same form as eq.~\eqref{eq:2ptoan}
with a 1-loop operator anomalous dimension~\footnote{ Let us notice that we have not used any special properties of the structure constants in this calculation, hence the above calculation is valid for any perturbed WZW model of the kind described in eq.~\eqref{eq:pwzw}, not necessarily conformal.
}
\begin{equation}\label{eq:andimaff}
\boldsymbol{\delta} = g  \bT+\mathcal{O}(g^2)\ .
\end{equation}
This formula is valid not just for the affine primary fields but also for all quasi-primary fields of the interacting theory. The computation involving the quasi-primary fields is similar, uses the OPEs of eq.~\eqref{eq:ope_quasi} and leads to the same result. In addition, let us remark that eq.~\eqref{eq:1laffcompact} gives the exact 1-loop correction to the 2-point correlation function for all quasi-primary fields.

The 2-loop computation of the correlation functions is slightly more involved. For convenience, let us define the shorthands
\beq
\label{eq:defDelta}
\mathcal{L}_i^a\colonequals \frac{L^a_{\Phi}}{z_i-u}+\frac{L^a_{\Psi}}{z_i-v}, \qquad \mathcal{R}_i^a\colonequals \frac{R^a_{\Phi}}{\bar{z}_i-\bar{u}}+\frac{R^a_{\Psi}}{\bar{z}_i-\bar{v}}.
\eeq
The Ward identities with two current insertions then read
\beq
\label{eq:twocurrentsinsertions}
\vac{J^a(z_1)J^b(z_2)\Phi(u,\bar{u})\Psi(v,\bar{v})}_0=
\left[\frac{k\eta^{ab}}{z_{12}^2}+
\frac{{f^{ab}}_c}{z_{12}}\mathcal{L}^c_2+
 (-1)^{|a||b|}\mathcal{L}_2^b\mathcal{L}_1^a\right] \cdot G^{(0)}(u,v)\ .
\eeq
Due to the symmetries of  structure constants, the above expression is graded-symmetric~\footnote{ Meaning that the exchange multiplies the correlation function by the sign factor $(-1)^{|a||b|}$.} under the simultaneous exchange $a\leftrightarrow b$ and $z_1\leftrightarrow z_2$.
Then, applying with care the global invariance condition of eq.~\eqref{eq:inv1} we get
\beq
\mathcal{L}^{b}_2\mathcal{L}^{a}_1\cdot G^{(0)}(u,v) = \left( h_1 h_2 L^{b}_\Phi L^{a}_\Phi +\frac{h_2}{z_1-v}{f^{ba}}_c L^c_\Phi\right)\cdot G^{(0)}(u,v)\ .
\eeq
Combining then the above with eq.~\eqref{eq:twocurrentsinsertions} and its antiholomorphic counterpart we obtain
\begin{multline}\label{eq:2loopint}
\vac{\Omega(z_1,\bar{z}_1)\Omega(z_2,\bar{z}_2)\Phi(u,\bar{u})\Psi(v,\bar{v})}_{0,c}=\left\{|h_1|^2|h_2|^2\bT^2+k\frac{h_1h_2}{\bar{z}_{12}^2}\Cas_L+k\frac{\bar{h}_1\bar{h}_2}{z_{12}^2}\Cas_R\right.\\
\left.- \left[
\frac{|u-v|^2}{2|z_{12}|^2(z_1-v)(\bar{z}_1-\bar{u})(z_2-u)(\bar{z}_2-\bar{v})}+ c.c.
\right]
c_{\mathrm{ad}} \bT
\right\}\cdot G^{(0)}(u,v)\ ,
\end{multline}
where $z_{ij}\colonequals z_i-z_j$ and $c_{\mathrm{ad}}$ is the eigenvalue of the Casimir in the adjoint representation, that is ${f^{bc}}_d{f^{ad}}_e\eta_{ab}=c_{\mathrm{ad}}\delta^c_e$.
It is a this point that the vanishing of the dual Coxeter number becomes important, since if $c_{\mathrm{ad}}\neq 0$,  the second line of eq.~\eqref{eq:2loopint} leads to divergences requiring the renormalization of the coupling constant. Assuming from now on that the Killing form is zero, we find that the  2-loop integrand simplifies to
\begin{equation}
\label{eq:loopcorrelator2}
\vac{\Omega(z_1,\bar{z}_1)\Omega(z_2,\bar{z}_2)\Phi(u,\bar{u})\Psi(v,\bar{v})}_{0,c}=\left(|h_1|^2|h_2|^2\bT^2+k\frac{h_1h_2}{\bar{z}_{12}^2}\Cas_L+k\frac{\bar{h}_1\bar{h}_2}{z_{12}^2}\Cas_R\right)\cdot G^{(0)}\ .
\end{equation}
Integrating it with the help of app.~\ref{sec:int}, we  obtain  the 2-loop correction
\beqa\label{eq:2loopcorr}
G^{(2)}=\tfrac{1}{2}g^2\left[\bT^2 \ell^2 +k (\Cas_L+\Cas_R) \ell\right]\cdot G^{(0)},
\eeqa
where we have defined
\beq
\label{eq:deffunctionell}
 \ell\colonequals2\log\tfrac{\epsilon^2}{|u-v|^2}.
\eeq

After this warm up, we can begin with the 3-loop calculation. First, we find that the Ward identities with three  current insertions read~\footnote{ For clarity of presentation, we have suppressed all signs coming from the gradation in eq.~\eqref{eq:3loopWardhol}.}
\begin{align}
\label{eq:3loopWardhol}
\vac{J^{a}_1J^{b}_2J^{c}_3\Phi\Psi}_{0,c}&=\left[ \frac{kf^{abc}}{z_{12}z_{13}z_{23}}+
\frac{kh_3}{z^2_{12}}\eta^{ab}L^{c}_\Phi
+\frac{kh_2}{z^2_{13}}\eta^{ac}L^{b}_\Phi
+\frac{kh_1}{z^2_{23}}\eta^{bc}L^{a}_\Phi\right.\nonumber\\ \notag
&+h_3 \left(  \frac{1}{z_{23}}-\frac{1}{z_2-v} \right)
\left(
\frac{{f^{ab}}_{\alpha}{f^{\alpha c}}_{\beta}}{z_{12}}+\frac{{f^{ac}}_{\alpha}{f^{b\alpha}}_{\beta}}{z_{13}}
+\frac{{f^{bc}}_{\alpha}{f^{\alpha a}}_{\beta}}{z_1-v}
\right)L^{\beta}_\Phi\\ \notag
&+ h_2h_3\left( \frac{1}{z_{12}}-\frac{1}{z_1-v} \right){f^{ab}}_{\alpha}L^{c}_\Phi L^{\alpha}_\Phi+
h_2h_3\left( \frac{1}{z_{13}}-\frac{1}{z_1-v} \right){f^{ac}}_{\alpha} L^{\alpha}_\Phi L^{b}_\Phi\\
&\left.+
h_1h_3\left( \frac{1}{z_{23}}-\frac{1}{z_2-v} \right){f^{bc}}_{\alpha} L^{\alpha}_\Phi L^{a}_\Phi
+ h_1h_2h_3 L^{c}_\Phi L^{b}_\Phi L^{a}_\Phi\right]\cdot G^{(0)}\ ,
\end{align}
where we have written $J^{a}_i:= J^{a}(z_i)$ and omitted the coordinate dependence on the primary fields $\Phi$ and $\Psi$.
Combining this expression with its antiholomorphic counterpart, we find
\begin{align}
\label{eq:3lintcc}
&\vac{\Omega_1(z_1,\bar{z}_1)\Omega(z_2,\bar{z}_2)\Omega(z_3,\bar{z}_3)\Phi(u,\bar{u})\Psi(v,\bar{v})}_{0,c}=\nonumber\\&=\left\{|h_1|^2|h_2|^2|h_3|^2\bT^3+
k\left[\left(|h_1|^2\frac{h_2h_3}{\bar{z}_{23}^2}+|h_2|^2\frac{h_1h_3}{\bar{z}_{13}^2}+|h_3|^2\frac{h_1h_2}{\bar{z}_{12}^2}\right)\Cas_L+\overline{(L\leftrightarrow R)}\right]\bT\right.\nonumber\\&+k^2\left(\frac{h_1\bar{h}_2}{\bar{z}_{13}^2z_{23}^2}+\frac{h_1\bar{h}_3}{\bar{z}_{12}^2z_{23}^2}+\frac{h_2\bar{h}_3}{\bar{z}_{12}^2z_{13}^2}+\mathrm{c.c.}\right)\bT +|h_3|^2\left[
|h_2|^2\left(\frac{1}{z_1-u}-\frac{1}{z_{12}}
\right)\left( \frac{1}{\bar{z}_{13}}-\frac{1}{\bar{z}_1-\bar{v}}
\right)\right.\nonumber\\& \left.\left. +
h_2\bar{h}_1\left(\frac{1}{z_{12}}-\frac{1}{z_{13}}
\right)\left(\frac{1}{\bar{z}_{23}}-\frac{1}{\bar{z}_2-\bar{v}}
\right)+\mathrm{c.c.}
\right] \bK \right\}\cdot G^{(0)}\ ,
\end{align}
where $\bK$ is a new invariant operator~\footnote{ The second equality can be proved as follows.
Evaluating the commutators and regrouping the structure constants one can rewrite
$([T_a,T_d],[T_b,T_c]) T^c T^d\otimes  [T^b,T^a] (-1)^{|b||d|} = - \str_{\mathrm{ad}}(T_e T_c T_d) T^c T^d \otimes T^e (-1)^{|c||d|}$.
The invariant rank 3 tensor appearing on the right hand side vanishes, see eq.~\eqref{eq:triangle}.}
\begin{align*}\label{eq:defK}
\bK\colonequals ([T_a,T_d],[T_b,T_c]) R^c_\Phi R^d_\Phi  L^b_\Phi    L^a_\Phi (-1)^{|b||d|}= ([T_a,T_d],[T_b,T_c]) R^c_\Phi   R^d_\Phi  L^a_\Phi  L^b_\Phi (-1)^{|b|(|a|+|d|)}\ .
\end{align*}
The above integrand is invariant under the exchange $z_i\leftrightarrow z_j$ and $u\leftrightarrow v$.
When reproducing eq.~\eqref{eq:3lintcc} it is useful to realize that only the last four terms of eq.~\eqref{eq:3loopWardhol} will contribute to $\bK$.

Let us comment a bit more on $\bK$. An essential difference between this new operator and the operators $\Cas_L$, $\Cas_R$ and $\bT$ appearing at lower orders is that it is built out of the structure constants. In that sense, it probes the non-abelian nature of the theory or, in other words, it ``feels'' the interaction. Indeed, the 1- and 2-loop integrands (\ref{eq:1laffcompact}, \ref{eq:loopcorrelator2}) look as if the perturbing currents were abelian.
The 3-loop integrand in eq.~\eqref{eq:3lintcc}, on the other hand, differs from an abelian current-current perturbation precisely by the terms proportional to $\bK$.

Integrating the terms in the first two lines of eq.~\eqref{eq:3lintcc} is as simple as
it is at 2-loops, see the formulas in app.~\ref{subsec:intmain}. The piece proportional
to $\bK$, however, is more complicated. The computation is done in the appendix
\ref{subsec:intstrange} and \emph{unexpectedly} the result is zero.
Thus, we have for the 3-loop correction
\beq\label{eq:3loopcor}
G^{(3)}=\tfrac{1}{6}g^3\left[ \bT^3\ell^3+3k\bT(\Cas_L+\Cas_R)\ell^2+6k^2 \bT\ell\right]\cdot G^{(0)}\ .
\eeq
Putting together eqs.~(\ref{eq:1laffcompact}, \ref{eq:2loopcorr}, \ref{eq:3loopcor}), discarding the higher powers of $\ell$ and comparing with eq.~\eqref{eq:2ptoan}
we read off the operator anomalous dimension of affine primaries up to 3-loops
\beq
\label{eq:3loopanomalousdimensions}
\boldsymbol{\delta} = g \bT+\tfrac{1}{2}g^2k(\Cas_L+\Cas_R)+g^3k^2 \bT+\calO(g^4)\ .
\eeq
Hence, the structure constants do not contribute to the anomalous dimensions of affine primaries, at least up to 3-loops.

At higher loops, other invariant operators  that are built out of the structure constants, will start contributing to the integrand.
Right now, we cannot compute these terms, let alone integrate them and check if they contribute to the anomalous dimension or not.

%%%%%%%%%%%%%%%%%%%%%%%%%%%%%%%%%%%%%%%%%%%%%%%%%%%%%%%%%%%%%%%
\subsection{Boundary affine primaries}\label{sec:bdth}
%%%%%%%%%%%%%%%%%%%%%%%%%%%%%%%%%%%%%%%%%%%%%%%%%%%%%%%%%%%%%%%

In this section, we would like to extend the bulk computations to the case when the model is defined on a  worldsheet with a boundary, which we take to be the upper half of the complex
plane $\mathbb{H}$.  We impose the following boundary conditions for the WZW currents:
\beq
\label{eq:boundaryconditions}
J^a(z)=\bar{J}^a(\bar{z})\ ,\qquad \text{ for } z=\bar{z}\ .
\eeq
Using the method of images, we make the identification $\bar{J}^a(\bar{z})=J^a(z^*)$, where $*$ indicates complex conjugation.
As in the bulk case, we concentrate on the computation of the 2-point functions of affine primary fields localized on the boundary.
Their defining OPEs are
\begin{equation}
\label{eq:OPEboundary}
J^a(z)\Phi(w)\sim \frac{B^a_{\Phi}\cdot\Phi(w) }{z-w}\ ,
\qquad \text{Im}(w)=0\ .
\end{equation}
There is no left or right action anymore, only a single one that we denote by $B$.

Let $\Phi$ and $\Psi$ be two primary fields.
Their 2-point function $G(u,v)$ can be computed as in eqs.~(\ref{eq:2pf}, \ref{eq:nG}), except that now the regularized domain
of integration at $n$-loops is
\beq
\label{eq:domainBn}
\calB_n\colonequals \big\{(z_1,\ldots,z_n)\in \mathbb{C}^n\ :\ |z_i-z^*_i|>\epsilon\ ,\; |z_i-z_j|>\epsilon\big\}\ .
\eeq
The operator anomalous dimension $\boldsymbol{\delta}$ can be extracted from the boundary equivalent of eq.~\eqref{eq:2ptoan}:
\begin{equation}
\label{eq:2ptoanboundary}
G(u,v) = \boldsymbol{a}(\boldsymbol{1} + \boldsymbol{\delta}\tilde{\ell}+\cdots )\cdot G^{(0)}(u,v)\ ,
\end{equation}
where $\tilde{\ell}=\log \tfrac{\epsilon^2}{|u-v|^2}$ is the appropriate boundary modification of eq.~\eqref{eq:deffunctionell}.

We can now compute the 1-loop correction with the help of the integrals in app.~\ref{subsec:intmain}
\begin{equation}
\label{eq:firstordercorrelationboundary}
G^{(1)}=-g\int_{\calB_1}\frac{d^2z_1}{\pi}|h_1|^2\eta_{ab}B_{\Phi}^aB_{\Phi}^b\cdot G^{(0)}= g\tilde{\ell} \Cas\cdot G^{(0)}\ ,
\end{equation}
where $\Cas\colonequals  \eta_{ab}B^a_\Phi B^b_\Phi$ and we have used the global invariance condition
\begin{equation}%\label{eq:}
(B^a_{\Phi}+B^a_\Psi)\cdot G^{(0)}(u,v)=0 \ .
\end{equation}
 At 2-loops, we start having current-current contractions appearing.
The integrand is computed as in the bulk case, with the difference that we have much more contractions now, since the perturbing field $\Omega$ contains two $J$ currents. All terms containing structure constants vanish and we obtain a result that looks as in an abelian theory,
\begin{multline}
\vac{\Omega(z_1,z_1^*)\Omega(z_2,z_2^*)\Phi(u)\Psi(v)}_{0,c}=\Big[|h_1|^2|h_2|^2\textbf{Cas}^2\nonumber\\+k\left(
\frac{h_1h_2}{(z_1^*-z_2^*)^2}+\frac{h_1^*h_2^*}{(z_1-z_2)^2}+\frac{h_1h_2^*}{(z_1^*-z_2)^2}+\frac{h_1^*h_2}{(z_1-z_2^*)^2}\right)\textbf{Cas}\Big]\cdot\vac{\Phi(u)\Psi(v)}_0.
\end{multline}
We can now use the recursion formulas \eqref{eq:intrecboundary} to perform the integration and obtain
\beq\label{eq:bint2}
G^{(2)}=g^2\left[\tfrac{1}{2}\tilde{\ell}^2\Cas^2+k\tilde{\ell}\Cas \right]\cdot G^{(0)}\ .
\eeq
As we shall argue in a moment, at 3-loops there are again no contributions involving structure constants.
We present only the final result
\beq
\label{eq:bint3}
G^{(3)}=g^3\left[\tfrac{1}{6}\tilde{\ell}^3\Cas^3+k\tilde{\ell}^2\Cas^2+k^2\tilde{\ell}\Cas \right]\cdot G^{(0)}\ .
\eeq
Putting together eqs.~(\ref{eq:firstordercorrelationboundary}, \ref{eq:bint2}, \ref{eq:bint3})
and comparing with eq.~\eqref{eq:2ptoanboundary}, we get the following expression for the operator anomalous dimension of boundary affine primaries at 3-loops
\beq
\label{eq:anomalousdimboundary}
\boldsymbol{\delta}=(g+kg^2+k^2 g^3)\Cas + \calO(g^4)\ .
\eeq
This result can be reproduced directly from~\cite{Schomerus:1999ug}, if we know beforehand that the structure constants do not contribute.

Let us now argue that the structure constants start contributing to the integrand at 4-loops.
We start by writing the $n$-loop correction in the following form
\beq
G^{(n)}(u,v)=\sum_{k=1}^{2n}t^{(n,k)}_{a_1\cdots a_k}(u,v)B_{\Phi}^{a_1}\cdots B_{\Phi}^{a_k}\cdot G^{(0)}(u,v),
\eeq
where $t^{(n,k)}_{a_1\cdots a_k}$ are invariant tensors constructed out of the structure constants ${f^{ab}}_{c}$ and the metric $\eta_{ab}$.
One can safely assume that $t^{(n,k)}_{a_1\cdots a_k}$ are completely (graded) symmetric, because otherwise one can
always reduce the number of $B$'s by using $[B_{\Phi}^a,B_{\Phi}^b]={f^{ab}}_cB_{\Phi}^c$  and then reabsorb the result in a lower rank tensor.
Second, notice that every term in $t^{(n,k)}_{a_1\cdots a_k}$ has at most $2n-k$ structure constants.
The product of $f$'s in every such term will be a symmetric tensor of rank at most $\min(k,2n-k)$, because a single structure constant contributes with  at most one index $a_i$.
Hence, if there are no invariant symmetric tensors of rank $\min(k,2n-k)$ or smaller built out of at most $2n-k$ structure constants, then the only contributions to $t^{(n,k)}_{a_1\cdots a_k}$ are made out of the $\eta_{ab}$. Clearly, there is nothing at rank 0 or 1 and due to the results of~\cite{Bershadsky:1999hk}, there is nothing at rank 2 either.
On the other hand, at 3-loops the only possibility is a tensor of rank 3 made out of 3 structure constants and the only such tensor is
$\text{str}_{\text{ad}}(T^{a_1}T^{a_2}T^{a_3})$.
The latter is indeed symmetric, implying that $\text{str}_{\text{ad}}([T^{a_1},T^{a_2}]T^{a_3})=0$, because the Killing form vanishes identically.
On the other hand it is also antisymmetric, since
\begin{equation}\label{eq:triangle}
\text{str}_{\text{ad}}(T^{a_1}T^{a_2}T^{a_3}) = ([T^{a_1},[T^{a_2},[T^{a_3},T_b]]],T^b) = -\varepsilon^{a_1a_2 a_3}([T^{a_3},[T^{a_2},[T^{a_1},T_b]]],T^b)\ ,
\end{equation}
and therefore must vanish identically. Here, we set $\varepsilon^{a_1a_2 a_3}\colonequals (-1)^{|a_1||a_2|+|a_1||a_3|+|a_2||a_3|}$ and borrowed some notation from app.~\ref{sec:conventions}. However, at 4-loops non-vanishing contributions from the structure constants to the integrand will appear, such as $\text{str}_{\text{ad}}(T^{a_1}T^{a_2}T^{a_3}T^{a_4})$ for instance.

%%%%%%%%%%%%%%%%%%%%%%%%%%%%%%%%%%%%%%%%%%%%%%%%%%%%%%%%%%%%%%%
%%%%%%%%%%%%%%%%%%%%%%%%%%%%%%%%%%%%%%%%%%%%%%%%%%%%%%%%%%%%%%%%%%%%
\section{Exact anomalous dimensions}
\label{sec:quasiabperttheory}
%%%%%%%%%%%%%%%%%%%%%%%%%%%%%%%%%%%%%%%%%%%%%%%%%%%%%%%%%%%%%%%
%%%%%%%%%%%%%%%%%%%%%%%%%%%%%%%%%%%%%%%%%%%%%%%%%%%%%%%%%%%%%%%

Our goal in this section is to single out a particular subsector
of fields for which anomalous dimensions may be computed to all
orders and to perform the relevant computations. To this end, we
carefully re-analyze a relevant observation made in
\cite{Bershadsky:1999hk} and deduce the precise conditions under
which the structure constants of the symmetry algebra may be dropped.
In the second subsection, we sum the entire perturbation series by treating
the perturbing fields as abelian.

%%%%%%%%%%%%%%%%%%%%%%%%%%%%%%%%%%%%%%%%%%%%%%%%%%%%%%%%%%%%%%%
\subsection{Maximally atypical fields}
\label{subsec:restrictionsonfields}
%%%%%%%%%%%%%%%%%%%%%%%%%%%%%%%%%%%%%%%%%%%%%%%%%%%%%%%%%%%%%%%

Consider two fields $\Phi,\Psi$ that have already  been  redefined to be quasi-primary in the interacting theory.
Suppose that $\Psi$ transforms in a representation $V$  with respect to the diagonal action $J^a_0 + \bar{J}^a_0$ of $\g$.
Then, the $n$-loop corrections to their perturbative 2-point function can  be brought to the form
\begin{equation}%\label{eq:}
 p(u,v) \vac{\Phi(u,\bar{u})\mathbf{E} \Psi(v,\bar{v})}_{0,c}\ ,
\end{equation}
where $\mathbf{E}\in \mathrm{End}_\mathfrak{g}V$ is an endomorphism commuting with the action of $\g$ and constructed out of the zero modes of the currents $J^a_0,\bar{J}^a_0$,
while $p(u,v)$ contains all the non-trivial loop integrals and must be a polynomial in $\ell$ of order at most $n$.
If the structure constants contribute to this correction, then
\begin{equation}\label{eq:E}
\mathbf{E} = {f^{ab}}_c {T^c}_{ba}(-1)^{|a|+|b|}\ ,
\end{equation}
where ${T^c}_{ba}\in\mathrm{End}V$ is an endomorphism which transforms covariantly with respect to the adjoint action
of $\mathfrak{g}$.
In other words, ${T^c}_{ba}=\varphi (T^c\otimes T_b\otimes T_a)$ for some  $\varphi\in\mathrm{Hom}_\mathfrak{g}(\mathfrak{g}^{\otimes 3},\mathrm{End}V)$, where $(T^a,T_c)=\delta^a_c$. We  would like to know when the structure  constants do not contribute to the 2-point function.

As pointed out in \cite{Bershadsky:1999hk}, this is the case when $\Psi$ is a scalar.
The proof is short. If $V$ is the trivial representation then ${T^c}_{ba}$ must be an invariant tensor.
In particular, if there is a representation $W$ of $\mathfrak{g}$ such that ${T^c}_{ba}$ is equal to the invariant tensor $ \str_W T^c T_bT_a$, or to any of its  permutations,
then $\mathbf{E}=0$.
Indeed, because  the Killing form vanishes identically one has, for instance
\begin{equation}%\label{eq:}
(-1)^{|a|+|b|} {f^{ab}}_c\,\str_W T^c T_bT_a =(-1)^{|a|+|b|}\tfrac{1}{2}{f^{ab}}_c\, \str_W T^c[T_b, T_a] = 0\ ,
\end{equation}
the same being true in all the other cases.
Then, using the fact that all invariant tensors are obtained
as linear combinations from supertraces of products
of generators in various orders and in various representations,
we conclude that $\mathbf{E}=0$ in general.

This proof generalizes to the case when $V$ is an irreducible representation of
$\mathfrak{g}$ with non-zero superdimension.
Indeed, taking the supertrace on both sides of eq.~\eqref{eq:E} and noticing that
$\str_V {T^c}_{ba}$ is an invariant tensor, it follows from the arguments we
outlined a moment ago that $\str_V \mathbf{E} =0$.
On the other hand, by the irreducibility of $V$, $\mathbf{E}$ is proportional
to the identity. Hence, $\mathbf{E}=0$ because we have assumed $\sdim V\neq 0$.

Notice that the condition $\sdim V\neq 0$ for an irreducible representation $V$
is a very strong one and, as far as we are aware, is known to hold only for the irreducible atypical representations of \emph{maximal} degree of atypicality.
In particular, the superdimensions of all typical representations vanish.

Suppose now that $V$ is an indecomposable representation which has an irreducible constituent (quotient of a submodule) of non-zero superdimension.
Then, a central endomorphism of $V$ will no longer be proportional to the identity in general, but can be split into a piece proportional to the identity and a nilpotent piece
$\mathbf{E} = \mathrm{E}\boldsymbol{1} + \mathbf{E}_n$.
Then, the above arguments only show that $\mathrm{E}=0$ if $\mathbf{E}$ is of the form~\eqref{eq:E}.

In conclusion, when computing the 2-point function~\eqref{eq:2ptoan} for a field
 $\Psi$ that transforms in an irreducible representation $V$  of $\g$ with $\sdim V\neq 0$,
we can safely drop all terms generated by the Ward identities~\eqref{eq:ward} as long as
they contain structure constants.
Moreover, if we do this for the 2-point function of a field $\Psi$ that transforms
in an indecomposable representation $V$, which is such that a subquotient of non-zero
superdimension exists, then only the anomalous dimension  $\delta$ and $a$ in
eq.~\eqref{eq:2ptoan} will be computed exactly. In other words, the structure
constants of our superalgebra can only contribute to the nilpotent pieces
$\boldsymbol{a}_n$ and $\boldsymbol{\delta}_n$. Let us stress that representation
theoretic arguments alone are not sufficient to determine $\delta$ and $a$
from the abelian approximation as soon as fields are not taken from the
described subsector of maximally atypical fields.

In the next section we shall sum up all contributions to the 2-point function
of quasi-primary fields that appear in the abelian approximation, that is when
we pretend that the perturbing currents are abelian. The resulting expression
must then be restricted to the maximally atypical subsector to obtain an exact
all loop formula for the anomalous dimensions of such fields.

%%%%%%%%%%%%%%%%%%%%%%%%%%%%%%%%%%%%%%%%%%%%%%%%%%%%%%%%%%%%%%%%%%%%
\subsection{Abelian perturbations in the bulk}
\label{subsec:abelpertbulk}
%%%%%%%%%%%%%%%%%%%%%%%%%%%%%%%%%%%%%%%%%%%%%%%%%%%%%%%%%%%%%%%%%%%%

Let $\Phi$ and $\Psi$ be two quasi-primary fields in the interacting theory.
We shall compute their 2-point function in the abelian approximation
\begin{equation}%\label{eq:}
J^a(z)J^b(w)\approx \frac{k \eta^{ab}}{(z-w)^2}\ ,\qquad
\bar{J}^a(\bar{z})\bar{J}^b(\bar{w})\approx \frac{k \eta^{ab}}{(\bar{z}-\bar{w})^2} \ .
\end{equation}
by using the Ward identities of eq.~\eqref{eq:ward} to evaluate the integrand in the expression~\eqref{eq:nG}.

\begin{figure}[ht]
\begin{center}
\includegraphics[width=12cm]{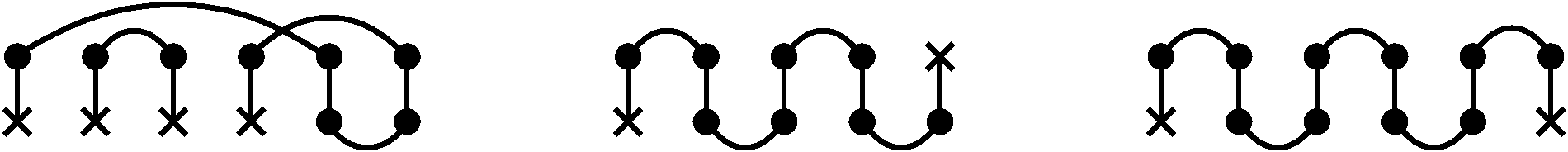}
\end{center}
\caption{Wick contractions appearing in the bulk perturbation theory. The middle diagram gives a contribution proportional to $\bT$, while the one on the right is proportional to $\Cas_L$.
\label{fig:diag1}
}
\end{figure}

Using Wick's theorem to normal order $n$ insertions of the interaction, we can write the $n$-loop integrand  as
\begin{equation}\label{eq:nloopfree}
\vac{ \Phi \Psi\prod_{i=1}^nJ^{a_i}_i\bar{J}^{b_i}_i\eta_{a_ib_i}}_{0,c} \approx \mathop{\mathop{\sum_{I,J\subset\{1,2,\cdots n\}}}_{|I|\equiv|J|\equiv n\text{ mod }2}} C^{IJ}_{a_I b_J} \vac{\Phi \Psi :\prod_{i\in I}J^{a_i}_i
\prod_{j\in J} \bar{J}^{b_j}_j :}_0\ ,
\end{equation}
where $C^{IJ}_{a_I b_J}$ is the correlation function of the remaining currents with vacuum diagrams removed, $a_I=\{a_i\}_{i\in I}$ and $b_J=\{b_j\}_{j\in J}$.
The integral of the summand labeled by $(I,J)$ will separate into exactly $\tfrac{1}{2}(|I|+|J|)$ non-factorisable integrals.
To see this one can write a diagrammatic expansion of $C^{IJ}_{a_Ib_J}$ into diagrams representing how the currents are contracted.
Denote the $i$-th insertion $J^{a_i}_i\bar{J}^{b_i}_i\eta_{a_ib_i}$ by a pair of connected vertices\, \includegraphics[height=3.5mm]{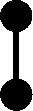}\, arranged vertically at the $i$-th position in a horizontal array of $n$ such pairs,
ordered from left to right, where
the vertex in the bottom corresponds to $J^{a_i}_i$ and the vertex in the top to $\bar{J}^{b_i}_i$.
We then cross out the uncontracted currents $\{J_i^{a_i}\}_{i\in I}$ and $\{\bar{J}_j^{b_j}\}_{j\in J}$.
A contraction between two currents will be represented by an edge \includegraphics[width=4mm]{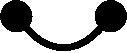} if it connects a pair of holomorphic currents and
an edge \includegraphics[width=4mm]{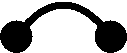} if they are antiholomorphic.
No horizontal edges are allowed to connect to crossed-out vertices.
Then, the Wick contractions in $C^{IJ}_{a_I b_J}$ can be formally written as a sum of diagrams connecting all the crosses  pairwise.
Examples of such diagrams are depicted in fig.~\ref{fig:diag1}.
Loops are not allowed because they correspond to vacuum diagrams.
Thus, the diagrams contributing to $C^{IJ}_{a_I b_J}$ will have exactly $\tfrac{1}{2}(|I|+|J|)$ connected pieces corresponding to the edges connecting crosses pairwise.
Clearly, every such edge gives rise to a non-factorisable integral.

Notice now that a disconnected  diagram will factorize into pieces that have already appeared at lower loops.
Hence, the new non-factorisable integrals at $n$-loops correspond to the diagrams with only two crossed currents.
Let us now count and integrate them. There are two cases to consider.

If $n$ is odd, then  one must cross out a holomorphic and an antiholomorphic current.
To compute the number of diagrams of this type one multiplies the number of choices made when drawing an edge starting at one cross
and passing through all other vertices until the second cross is reached.
Thus, there are $(n-2)!$ ways to connect the crosses when their position is fixed. After summing over all positions one gets a total of $n!$ diagrams.
Consider, for example, the contribution to eq.~\eqref{eq:nloopfree} represented by the diagram in the middle of fig.~\ref{fig:diag1}
\begin{equation}\label{eq:el_int_odd}
\int_{\mathcal{D}_n} \frac{d^2z_1\cdots d^2z_{n}}{\pi^{n}} \vac{\Phi \Psi J^a_1\bar{J}^b_n\eta_{ab}} \mathop{\prod^{n-2}_{i=1}}_{i \text{ odd}}\frac{k^2}{\bar{z}^2_{i,i+1}z^2_{i+1,i+2}}\ .
\end{equation}
In order to perform the integrals we need to evaluate the remaining correlator.
With the OPEs
\begin{equation}\label{eq:ope_quasi}
J^a(z)\Phi(w,\bar{w})\sim \sum_{n=0}^\infty\frac{(J^a_n\Phi)(w,\bar{w})}{(z-w)^{n+1}}\ ,\qquad \bar{J}^a(\bar{z})\Phi(w,\bar{w})\sim \sum_{n=0}^\infty\frac{(\bar{J}^a_n\Phi)(w,\bar{w})}{(\bar{z}-\bar{w})^{n+1}}\ ,
\end{equation}
the similar OPEs for $\Psi$ and the Ward identities of eq.~\eqref{eq:ward} we can integrate over $z_1$ and $z_n$ with the help of the elementary
integrals  (\ref{eq:integralcalJ12}, \ref{eq:intrecbulk}) from app.~\ref{subsec:intmain}.
Notice that only the first order poles in the OPEs of formula~\eqref{eq:ope_quasi} will contribute to these two integrations, after which eq.~\eqref{eq:el_int_odd} becomes
\begin{equation}\label{eq:fin_int_odd}
k^2 \int_{\mathcal{D}_n} \frac{d^2z_2\cdots d^2z_{n-1}}{\pi^{n-2}} \bar{h}_2h_{n-1}\left(\ \mathop{\prod^{n-3}_{i=2}}_{i \text{ even}}\frac{k^2}{z^2_{i,i+1}\bar{z}^2_{i+1,i+2}}\right) \bT \cdot G^{(0)} = -k^{n-1} \ell\ \bT \cdot G^{(0)}
\end{equation}
where we have used eqs.~\eqref{eq:intrecbulk} to perform the remaining integration over $z_2,\dots, z_{n-2}$ successively and closed the chain with the basic integral~\eqref{eq:basicintegral1}. All other diagrams with a cross on top and a cross in the bottom will
contribute exactly with the same amount as eq.~\eqref{eq:fin_int_odd}, because the corresponding integrals can be identified with eq.~\eqref{eq:el_int_odd} after a relabeling of indices.

If $n$ is even then one must cross out either two holomorphic or two antiholomorphic currents.
For a fixed position of crosses there are again $(n-2)!$ diagrams.
Thus, summing over positions we get a total of $\tfrac{1}{2}n!$ diagrams with the crosses located either in the bottom or on top.
The contribution to eq.~\eqref{eq:nloopfree} of a sample diagram represented on the right of fig.~\ref{fig:diag1} will be
\begin{equation}\label{eq:el_int_ev}
\int_{\mathcal{D}_n} \frac{d^2z_1\cdots d^2z_{n}}{\pi^{n}} \vac{\Phi \Psi J^a_1J^b_n\eta_{ab}} \frac{k}{\bar{z}^2_{n-1,n}}\mathop{\prod^{n-3}_{i=1}}_{i \text{ odd}}\frac{k^2}{\bar{z}^2_{i,i+1}z^2_{i+1,i+2}} = k^{n-1}\ell\ \Cas_L \cdot\ G^{(0)}\ ,
\end{equation}
where the integration was done in the same way as above.
All  other diagrams with two crosses in the bottom contribute with the same amount as formula~\eqref{eq:el_int_ev}.
Also, by the same arguments, the contribution of every diagram with two crosses in the top will be
\begin{equation}\label{eq:el_int_ev1}
 k^{n-1}\ell\ \Cas_R \cdot\ G^{(0)}\ .
\end{equation}

Our first conclusion  is  that all connected diagrams at $n$-loops are proportional to $\ell$ after integration.
Hence diagrams that factorize into $m=\tfrac{1}{2}(|I|+|J|)$ connected pieces will be proportional to $\ell^{m}$.
Our  second conclusion is that the non-factorisable diagrams that we have just computed are precisely the ones that contribute to the first two terms in eq.~\eqref{eq:2ptoan}.

Putting together eqs.~(\ref{eq:nG}, \ref{eq:fin_int_odd}, \ref{eq:el_int_ev}, \ref{eq:el_int_ev1})
we thus get for the following all loop correction
\begin{align}%\label{eq:}
G^{(2m-1)} &\approx k^{2m-2} g^{2m-1}\ell \ \bT\cdot G^{(0)}+\cdots\ ,\nonumber\\
G^{(2m)} &\approx \tfrac{1}{2}k^{2m-1}g^{2m}\ell\ (\Cas_L+\Cas_R)\cdot G^{(0)}+\cdots\ ,
\end{align}
where the dots denote higher powers of $\ell$.
Summing up the geometric series, we arrive at a compact result for the operator
anomalous dimensions \footnote{A similar expression has been obtained by somewhat
different techniques in \cite{Konechny}.}
\beq
\label{eq:anomalousdimensionsallloop}
\boldsymbol{\delta}\approx\frac{1}{1-g^2k^2}\big[g \bT+\tfrac{1}{2}kg^2(\Cas_L+\Cas_R)\big]\ ,
\eeq
in the abelian approximation. We also have $\boldsymbol{a}\approx \boldsymbol{1}$.
The conditions under which the abelian approximation provides exact expressions for
the operators $\boldsymbol{\delta}$ and $\boldsymbol{a}$  or for their diagonal
contributions $\delta$ and $a$ were spelled out at the end of
sec.~\ref{subsec:restrictionsonfields}. We will specialize the general formula
\eqref{eq:anomalousdimensionsallloop} to these cases in the concluding section.

Obviously, our calculation must be exact for abelian perturbations.
The simplest example of an abelian perturbation is the massless Thirring model or, after bosonization, the  compactified free boson.
The role of the WZW model is played by a complex free fermion.
The abelian perturbation changes the radius $r$ of the compact boson,  with $r=1$ corresponding to the unperturbed point.
The conformal dimensions as a function of the radius $r$ read
\beq
\label{eq:partitionfunctionfreeboson}
h_{m,w}(r)=\frac{1}{2}\left(\frac{m}{2r}+wr\right)^2, \quad  \bar{h}_{m,w}(r)=\frac{1}{2}\left(\frac{m}{2r}-wr\right)^2
\eeq
with either $m\in 2\mathbb{Z}$ and $w\in \mathbb{Z}$  or   $m\in2\mathbb{Z}+1$ and $w\in\mathbb{Z}+\tfrac{1}{2}$.
The  free fermions at $r=1$ correspond to $(m,w)=(\pm 1,\pm \tfrac{1}{2})$ and $(m,w)=(\mp 1,\pm \tfrac{1}{2})$.
If the $\BU{1}\times \BU{1}$ abelian currents are normalized as  $J(z)J(w)\sim (z-w)^{-2}$, $\bar{J}(\bar{z})\bar{J}(\bar{w})\sim (\bar{z}-\bar{w})^{-2}$, then
$k=1$ and the left, respectively right $\BU{1}$ charges are $\frac{m}{2}+w$, respectively  $\frac{m}{2}-w$.
The quadratic Casimir is simply the square of the charge so that we have
\beq
\bT(m,w)=\frac{m^2}{4}-w^2, \qquad \tfrac{1}{2}\left(\Cas_{L}+\Cas_{R}\right)(m,w)=\frac{m^2}{4}+w^2.
\eeq
Plugging this into the formula for the conformal dimensions coming from eq.~\eqref{eq:partitionfunctionfreeboson} we get an  agreement with eq.~\eqref{eq:anomalousdimensionsallloop}
\beq\label{eq:anom_thirr}
\delta=h_{m,w}(r)-h_{m,w}(1)=\frac{1}{1-g^2}\left[g\left(\frac{m^2}{4}-w^2\right)+g^2\left(\frac{m^2}{4}+w^2\right)\right],
\eeq
if the compactification radius $r$ is related to the coupling $g$ via the equation $r^2=\frac{1-g}{1+g}$.
This relation can be computed directly by carefully bosonizing the Thirring model, see \cite{Klassen:1992eq}.

%%%%%%%%%%%%%%%%%%%%%%%%%%%%%%%%%%%%%%%%%%%%%%%%%%%%%%%%%%%%%%%%%%%%
%%%%%%%%%%%%%%%%%%%%%%%%%%%%%%%%%%%%%%%%%%%%%%%%%%%%%%%%%%%%%%%%%%%%

\section{Discussion and Outlook}

%%%%%%%%%%%%%%%%%%%%%%%%%%%%%%%%%%%%%%%%%%%%%%%%%%%%%%%%%%%%%%%%%%%%
%%%%%%%%%%%%%%%%%%%%%%%%%%%%%%%%%%%%%%%%%%%%%%%%%%%%%%%%%%%%%%%%%%%%

In this paper we have studied 2-point functions in current-current
deformations of WZW models for Ricci-flat (simple) supergroups $\G$. Special
emphasis was put on the computation of anomalous dimensions. For
affine primaries, we were able to compute the full 2-point function
up to 3-loops. Quite remarkably, all non-vanishing contributions
were quasi-abelian, i.e.\ while terms containing the structure
constants of the Lie superalgebra did show up, at least in the bulk,
they appeared but in front of vanishing loop integrals. For
the conformal weights of quasi-primary fields in maximally atypical
representations of the (diagonal) $\G$ action we argued that the
perturbative expansion does not receive contributions from the structure constants.
In addition we summed the quasi-abelian
perturbative expansion to obtain exact all-loop expressions for the
anomalous dimensions of such fields.

Maximally atypical fields are certainly a very small subset of the
entire space of bulk fields. But they still contain very valuable
information about the CFT.
The sector these
fields span may be referred to as $\tfrac{1}{2}$BPS, the difference being only that the relevant
symmetry action comes from an internal symmetry $\G$ rather than
one acting on space-time. Matching $\tfrac{1}{2}$BPS sectors of two models
can be a powerful first approach to dualities, such as the duality
between Calabi-Yau compactifications and Gepner models or between
$\mathcal{N}=4$ super Yang-Mills theory and strings in $AdS_5 \times S^5$. In the context of current-current deformed WZW models, the spectrum of the world-sheet dilation operator in the $\tfrac{1}{2}$BPS like sector of the target space symmetry was shown to obey Casimir evolution so that it can be tracked very easily. The only input we need is the spectrum of fields at the WZW point.

%%%%%%%%%%%%%%%%%%%%%%%%%%%%%%%%%%%%%%%%%%%%
Note that there are many ways to build $\tfrac{1}{2}$BPS fields for the diagonal action from ``combinations'' of left- and right-moving fields at the WZW-point.
Indeed, a  field $\Phi$ of the deformed theory which transforms in a maximally atypical representation
$V_D$ of the diagonal symmetry group $\G$ will be at the WZW point part of a bigger representation  $V_L\otimes V_R$ of the enhanced  global symmetry group $\G\times \G$.
Technically speaking,  $\Phi$ is constructed by applying an intertwiner from the WZW multiplet $V_L\otimes V_R$  to $V_D$.
Atypical representations can be grouped into ``blocks'' in such a way that the quadratic (and higher order) Casimirs have the same eigenvalues in all representations of a block, see~\cite{Serganova:1998}.
For $\mathfrak{psl}(n|n)$ and $D(2,1;\alpha)$ there is only one block of maximally atypical representations --- the one  containing the trivial representation. In this block~\footnote{ Here, we denote by $\mathrm{Cas}$ the eigenvalues of the appropriate quadratic Casimir operators $\Cas$. } $\mathrm{Cas}_D=0$  and the formula for the anomalous dimensions of eq.~\eqref{eq:anomalousdimensionsallloop} simplifies
\begin{equation}%\label{eq:}
\mathfrak{psl}(n|n)\ ,\ D(2,1;\alpha):\qquad \delta_0 = -\frac{g}{2(1+gk)}(\mathrm{Cas}_L+\mathrm{Cas}_R)
\end{equation}
 if we use eq.~\eqref{eq:definitionofT}.
On the other hand,  the maximally atypical blocks of $\mathfrak{osp}(2n+2|2n)$ are labeled by an integer $m\geq 0$.
The eigenvalue of the Casimir in the $m$-th block is $\mathrm{Cas}_D= m^2$ and the anomalous dimension~\eqref{eq:anomalousdimensionsallloop} simplifies to
\begin{equation}\label{eq:maxatyp_osp}
\mathfrak{osp}(2n+2|2n):\qquad \delta_m = \frac{g m^2}{2(1-g^2k^2)}-\frac{g}{2(1+gk)}(\mathrm{Cas}_L+\mathrm{Cas}_R)\ .
\end{equation}
%

%%%%%%%%%%%%%%%%%%%%%%%%%%%%%%%%%%%%
Note that with respect to the left and right action of the unperturbed model, the $\tfrac{1}{2}$BPS states for the diagonal action can transform in typical or atypical representations, i.e.\ belong to long
or short multiplets. Because of its rich structure, the $\tfrac{1}{2}$BPS sector
of deformed WZW models is like a footprint. Two models with the same
$\tfrac{1}{2}$BPS sector have a good chance to be dual descriptions of each other. Therefore, our results should provide a useful new tool in discovering
non-perturbative world-sheet dualities.

There are a number of well-known theories that appear as
current-current deformations of a WZW model. One particularly interesting series is given by the $\osp{2n+2}{2n}$ Gross-Neveu models, see \cite{Candu:2010yg} for details on the notation. When $n=0$, the associated Gross-Neveu model is the massless Thirring model.
Note that in this case eq.~\eqref{eq:maxatyp_osp} reproduces the entire spectrum~\eqref{eq:anom_thirr}.
For $n\geq 1$, the Gross-Neveu model becomes a complicated interacting theory. Its holomorphic field content consists of $2n+2$ real fermions  and $n$ $\beta\gamma$-systems, all of which have conformal dimension $h=\frac{1}{2}$ when the coupling $g$ is turned off. These fields are grouped in a single vector $\Psi$ that transforms in the fundamental representation of $\osp{2n+2}{2n}$. The action reads
\beq
\calS_{\text{GN}}=  \int  \frac{d^2z}{2\pi} \left[ \Psi\cdot
\bar{\partial}\Psi+ \bar{\Psi}
\cdot \partial\bar{\Psi}+g  (\Psi \cdot \bar{\Psi})^2\right]\ ,
\eeq
where $\cdot$ is the invariant scalar product in the fundamental representation.
Because of the vanishing of the dual Coxeter number, the theory is
conformal with central charge given by $c=1$. For $g=0$, the
Gross-Neveu model can be understood as an $\osp{2n+2}{2n}$ WZW
model at level $k=1$ and the interaction term as a current-current
perturbation. Hence, $\osp{2n+2}{2n}$ Gross-Neveu models provide
a whole series of examples to which the results of this work
can be applied.

There are a number of obvious open problems to be addressed. First,
it would be very interesting to examine the proposed duality between
\osp{2n+2}{2n} Gross-Neveu models and sigma models on odd-dimensional
superspheres $S^{2n+1|2n}$ through a
comparison of the maximal atypical bulk spectrum. We believe that such an analysis will resemble the successful test performed in \cite{Mitev:2008yt} where the entire spectrum of boundary fields for the unique maximally symmetric boundary condition has been matched. Secondly, once the maximally atypical sector of the superspheres is under good control, one could return to the sigma models on complex projective superspaces that were studied in \cite{Candu:2009ep}. These models are believed to be dual to \psl{n} WZW models, see \cite{Candu:2011hu} and further references therein, though not much supporting evidence has been provided. In particular, no WZW-point could be identified in the boundary spectra for the sigma model on $\mathbb{CP}^{1|2}$ that were found in \cite{Candu:2009ep}.
In the case of complex projective superspaces, the study of boundary spectra is hardly conclusive since there exists an infinite family of maximally symmetric boundary conditions and hence there is considerable freedom in matching the spectra. The maximally atypical subsectors in the bulk theory, however, are unique and therefore our new results could bring a conclusion about the conjectured dual of sigma models on complex projective superspaces within reach. Of course, it might also be worthwhile to investigate the spectra of other deformed WZW models to look for non-perturbative dualities that have not been conjectured before.

In the introduction we mentioned another very important class of CFTs
with internal supersymmetry: conformal sigma models on symmetric superspaces $\G/\,\mathrm{H}$ classified in~\cite{Candu:2010yg}. These will be treated in a companion article, in which we intend to present the computation of the 2-point functions for all fields to 1-loop. We have also considered models with world-sheet supersymmetry. The Lagrangian of such models contains terms similar to the Gross-Neveu model interaction whose contribution to the anomalous dimensions has been computed above. We are currently in the process of investigating supersymmetric sigma models in detail and expect to find,  at the very least,  a 1-loop expression for the anomalous dimensions. In the context of WZW models, it might also be interesting to analyze those exactly marginal deformations described which preserve the left and the right global symmetries separately. Continuing along these lines, let us note that both conformal sigma models and Gross-Neveu models are special members of a much larger class of CFTs, namely those with continuously varying exponents. As shown in \cite{Candu:2011hu}, there exists six different families of superspace GKO coset models that possess exactly marginal deformations. These perturbations might also be amenable to an exact perturbative treatment, similar to the one we have performed in this paper.

In a somewhat different direction, it might also be worthwhile pushing our exact perturbative computations of 2-point functions for affine primaries to higher orders and to look for the first signatures of the non-abelian nature of the symmetry $\G$. Recall that such terms are allowed by group theory but did not appear in our 3-loop computation because the associated integral vanishes. The restriction to quasi-abelian contributions in the perturbative expansion worked very well. It would be very interesting to explain this amazing success of quasi-abelian perturbation theory in boundary spectra with analytic means.

%%%%%%%%%%%%%%%%%%%%%%%%%%%%%%%%%%%%%%%%%%%%%%%%%%%%%%%%%%%%%%%
\section*{Acknowledgments}
The authors wish to thank Anatoly Konechny and Thomas Quella for discussions on the quasi-abelian terms. We also thank Burkhard Eden, Simeon Hellerman and Christoph Sieg for their interest, advice and critique. This work was supported in part by the SFB 676, project A9.

%%%%%%%%%%%%%%%%%%%%%%%%%%%%%%%%%%%%%%%%%%%%%%%%%%%%%%%%%%%%%%%
\appendix
%%%%%%%%%%%%%%%%%%%%%%%%%%%%%%%%%%%%%%%%%%%%%%%%%%%%%%%%%%%%%%%

%%%%%%%%%%%%%%%%%%%%%%%%%%%%%%%%%%%%%%%%%%%%%%%%%%%%%%%%%%%%%%%
\section{Signs and conventions}
\label{sec:conventions}
%%%%%%%%%%%%%%%%%%%%%%%%%%%%%%%%%%%%%%%%%%%%%%%%%%%%%%%%%%%%%%%

We denote the basis of the superalgebra $\fg$ by $(T^a)_{a=1}^{\dim \fg}$.
Every basis element $T^a$ has a well defined degree $|a|:=|T^a|$, which is 0 or 1 depending on whether $T^a$ it is bosonic or fermionic.
The structure constants are defined by $[T^a,T^b]= {f^{ab}}_c T^c$.
The superalgebra $\fg$  has an invariant, non-degenerate, consistent, graded-symmetric bilinear form which is denoted by $(\, , )$ and whose matrix elements in the above basis we write as $\eta^{ab}:=(T^a,T^b)$.
The dual basis $(T_a)_{a=1}^{\dim \fg}$ is defined as $(T^a,T_b)=\delta^a_b$ (the order is important). Explicitly, one has $T_a=T^b\eta_{ba}$, where $\eta^{ab}\eta_{bc}=\delta^a_c$.
Hence, $(T_a,T_b)=\eta_{ba}$ (notice the order).
With the above conventions ${f^{ab}}_c = ([T^a,T^b],T_c)$ and we set $f^{abc}:= ([T^a,T^b],T^c)$.

The currents of a WZW model at level $k$ can be written in terms of a supergroup valued map $\tg:\Sigma\rightarrow \G$ as
\beq
J\colonequals -k\partial \tg \tg^{-1}, \qquad \bar{J}\colonequals k\tg^{-1}\bar{\partial} \tg\ .
\eeq
They are  \textit{even} objects because the group element $\tg$ is even.
We can write them in components as $J=T_aJ^a$, $\bar{J}=T_a\bar{J}^a$, where again the order is important.
This is because $T_a$ can be identified, in the standard way, with a tangent vector at identity
and will therefore only commute with $J^a$ in the graded sense.
Now, our perturbing field can be written as $\Omega\colonequals (J,\bar{J})=(\bar{J},J)$ or, in components, as
\beq
\label{eq:defOmega}
\Omega=(J,\bar{J})=(T_a,\bar{J})J^a=(T_a,T_b)\bar{J}^bJ^a=\eta_{ba}\bar{J}^bJ^a=\eta_{ab}J^a\bar{J}^b\ .
\eeq

%%%%%%%%%%%%%%%%%%%%%%%%%%%%%%%%%%%%%%%%%%%%%%%%%%%%%%%%%%%%%%%
\section{Integrals}
\label{sec:int}
%%%%%%%%%%%%%%%%%%%%%%%%%%%%%%%%%%%%%%%%%%%%%%%%%%%%%%%%%%%%%%%

In this appendix, we present the detailed computation of the various integrals that appear throughout the main text.

%%%%%%%%%%%%%%%%%%%%%%%%%%%%%%%%%%%%%%%%%%%%%%%%%%%%%%%%%%%%%%%
\subsection{Main integral formulas}
\label{subsec:intmain}
%%%%%%%%%%%%%%%%%%%%%%%%%%%%%%%%%%%%%%%%%%%%%%%%%%%%%%%%%%%%%%%

\begin{figure}[h]
\begin{center}
\includegraphics[trim = 210mm 210mm 205mm 205mm, clip, width=6.5cm]{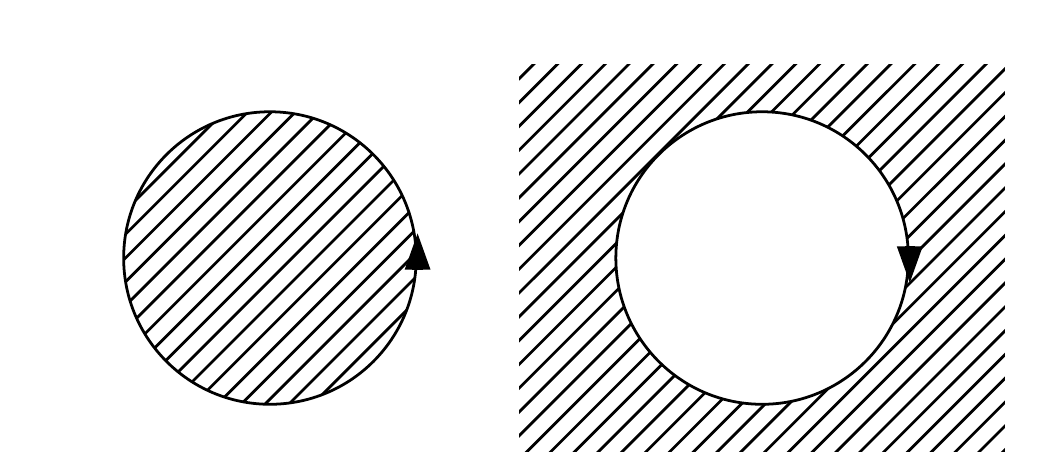}
\end{center}
\caption{The direction of the contour depends on the domain $M$, which is indicated here by the filled region. If the normal vector of the domain $M$ point outwards, then the contour is counterclockwise. On the other hand, if it points inwards, then it is clockwise.}
\label{fig:contourdirection}
\end{figure}

In our notation, if $z=x+iy$, then $d^2z=dxdy$. All the bulk integrals are performed over the regularized domains $\calD_n$ of equation \eqref{eq:defdomainD}, while the boundary ones use $\calB_n$ defined in eq.~\eqref{eq:domainBn}. The computation of the integrals is done using Stokes' theorem, which in complex coordinates reads
\beq
\label{eq:Stokes2d}
\int_M d^2z\,(\partial A+\bar{\partial}\bar{A})=\frac{i}{2}\oint_{\partial M}(d\bar{z}A-dz\bar{A}),
\eeq
where the contour integral is performed as depicted in figure \ref{fig:contourdirection}.

The first integral we wish to compute appeared in eq.~\eqref{eq:1laffcompact} and reads
\beq
\label{eq:basicintegral0}
\mathcal{I}\colonequals \int_{\calD_1}\frac{d^2z}{\pi} \left(\frac{1}{z-u}-\frac{1}{z-v}\right)\left(\frac{1}{\bar{z}-\bar{u}}-\frac{1}{\bar{z}-\bar{v}}\right)=\oint_{\partial\calD_1}\frac{dz}{2\pi i}\left(\frac{1}{z-u}-\frac{1}{z-v}\right)\log\left|\frac{z-u}{z-v}\right|^2,
\eeq
where $u$ and $v$ are any complex numbers with $u\neq v$ and we have made use of eq.~\eqref{eq:Stokes2d} in such a way as to have a well defined integrand in the whole complex plane. The contour integration contains a piece around $u$ and another around $v$, both of which give an equal contribution, leading to
\beq
\label{eq:basicintegral1}
\mathcal{I}=2\log\left(\frac{|u-v|}{\epsilon}\right)^2 +\calO(\epsilon^2)=-\ell+\calO(\epsilon^2),
\eeq
where we remind that the function $\ell$ was defined back in eq.~\eqref{eq:deffunctionell}. For the computation of the 3-loop bulk integral of section \ref{subsec:intstrange} we need a straightforward generalization of the basic integral of eq.~\eqref{eq:basicintegral0}, namely
\beq
\label{eq:basicintegral3}
\int_{\calD_1}\frac{d^2z}{\pi}\left(\frac{1}{z-x_1}-\frac{1}{z-x_2}\right)\left(\frac{1}{\bar{z}-\bar{x}_3}-\frac{1}{\bar{z}-\bar{x}_4}\right)=\log\frac{|x_{14}|^2|x_{23}|^2}{|x_{13}|^2|x_{24}|^2}.
\eeq
The second kind of integrals we need to compute appear at 2-loops and consists of
\beq
\mathcal{J}_{m,n}\colonequals  \int_{\calD_1} \frac{d^2z}{\pi}\frac{1}{(z-u)^{m+1}(\bar{z}-\bar{v})^{n+1}},
\eeq
where we require for convergence's sake that $m+n\geq 1$. If $m\geq 1$, we can write the integrand of $\mathcal{J}$ as a well defined holomorphic derivative, while if $n\geq 1$ we can write it as an antiholomorphic derivative. A simple computation leads to the succinct expression
\beq
\label{eq:integralcalJ12}
\calJ_{m,n}=\frac{(-1)^m\delta_{n,0}}{m(u-v)^m}+\frac{\delta_{m,0}}{n(\bar{u}-\bar{v})^n}+\calO(\epsilon).
\eeq
Unlike our other integrals, these remain finite when the regulator is set to zero. Using equation \eqref{eq:integralcalJ12}, one finds that we can obtain the following useful recursion relations
\beq
\label{eq:intrecbulk}
\int_{\calD_1}\frac{d^2z_1}{\pi}\frac{h_1}{\bar{z}_{12}^2}=-\bar{h}_2, \qquad \int_{\calD_1}\frac{d^2z_1}{\pi}\frac{\bar{h}_1}{z_{12}^2}=-h_2.
\eeq

In the boundary theory, the recursion formulas  \eqref{eq:intrecbulk} need to be modified. One finds using contour integral techniques, that
\beq
\label{eq:intrecboundary}
\int_{\calB_1}\frac{d^2z_1}{\pi}\frac{h_1}{\bar{z}_{12}^2}=\int_{\calB_1}\frac{d^2z_1}{\pi}\frac{\bar{h}_1}{z_{12}^2}=0,\quad \int_{\calB_1}\frac{d^2z_1}{\pi}\frac{h_1}{(\bar{z}_1-z_2)^2}=-h_2,\quad \int_{\calB_1}\frac{d^2z_1}{\pi}\frac{\bar{h}_1}{(z_1-\bar{z}_2)^2}=-\bar{h}_2.
\eeq
Proving the above by using eq.~\eqref{eq:Stokes2d} is easy. For instance one finds
\begin{multline}
\int_{\calB_1}\frac{d^2z_1}{\pi}\frac{h_1}{(\bar{z}_1-z_2)^2}=-\oint_{\partial \calB_1}\frac{dz_1}{2\pi i}\frac{h_1}{(\bar{z}_1-z_2)}\\=-\int_{-\infty}^{\infty}\frac{dx}{2\pi i}\left(\frac{1}{x-u+i\epsilon}-\frac{1}{x-v+i\epsilon}\right)\frac{1}{x-z_2-i\epsilon}+\calO(\epsilon)=-h_2+\calO(\epsilon)
\end{multline}
and similarly for the other integrals.

%%%%%%%%%%%%%%%%%%%%%%%%%%%%%%%%%%%%%%%%%%%%%%%%%%%%%%%%%%%%%%%
\subsection{Computing the 3-loop term}
\label{subsec:intstrange}
%%%%%%%%%%%%%%%%%%%%%%%%%%%%%%%%%%%%%%%%%%%%%%%%%%%%%%%%%%%%%%%

\begin{figure}[h]
\begin{center}
\includegraphics[trim = 210mm 210mm 205mm 205mm, clip, width=9cm]{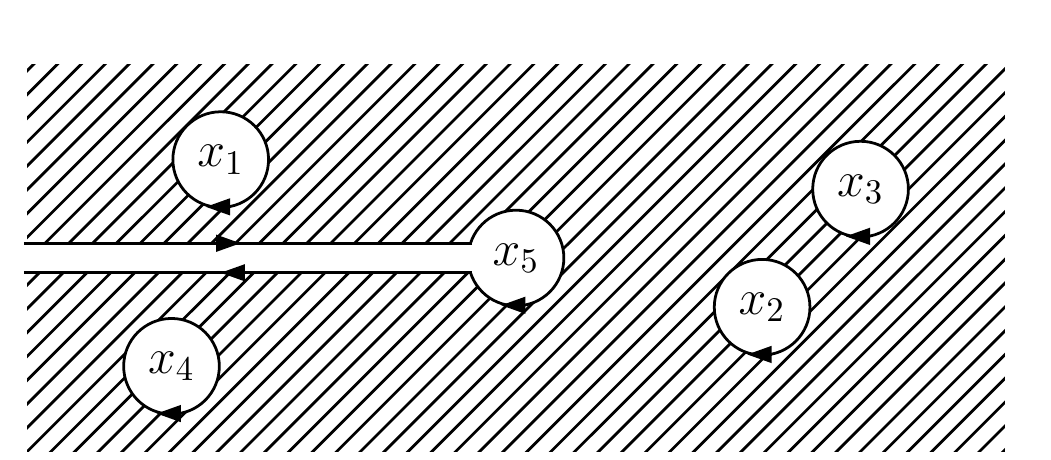}
\end{center}
\caption{The integral contour for $\calH$. There are no contributions coming from spatial infinity.}
\label{fig:contourintegral}
\end{figure}

In order to calculate the integral of the term appearing in from of $\bK$ in \eqref{eq:3lintcc}, we now have to look at a special class of integrals containing logarithms. Specifically, we define
\beq
\label{eq:intlog1}
\mathcal{H}^{x_1,x_2,x_3,x_4}_{x_5}\colonequals \int_{\calD}\frac{d^2z}{\pi}2\text{Re}\left[\left(\frac{1}{z-x_1}-\frac{1}{z-x_2}\right)\left(\frac{1}{\bar{z}-\bar{x}_3}-\frac{1}{\bar{z}-\bar{x}_4}\right)\right]\log|z-x_5|^2.
\eeq
The domain $\calD$ is defined by first excluding small disks of radius $\epsilon$ around the points $x_i$ and then by cutting  a small horizontal strip starting at $x_5$ and stretching to $x_5-\infty$. Its precise shape and the orientation of its boundary is depicted in figure \ref{fig:contourintegral}. In the limit in which the width of the strip goes to zero, we recover our usual integration domain. 
By using the identity $\log|z-x|^2=\log(z-x)+\log(\bar{z}-\bar{x})$ and setting $k_{xy}\colonequals \frac{1}{z-x}-\frac{1}{z-y}$, we can write
\begin{multline}
\mathcal{H}= \int_{\calD}\frac{d^2z}{\pi}\left\{\partial\left[\left(\log\left|\frac{z-x_1}{z-x_2}\right|^2\bar{k}_{x_3x_4}+\log\left|\frac{z-x_3}{z-x_4}\right|^2\bar{k}_{x_1x_2}\right)\log(\bar{z}-\bar{x}_5)\right]\right.\\\left.+\bar{\partial}\left[\left(\log\left|\frac{z-x_1}{z-x_2}\right|^2k_{x_3x_4}+\log\left|\frac{z-x_3}{z-x_4}\right|^2k_{x_1x_2}\right)\log(z-x_5)\right]\right\}.
\end{multline}
Using Stokes' theorem, we find
\begin{multline}
\mathcal{H}=\oint_{\partial\calD}\left\{\frac{d\bar{z}}{2\pi i}\left[\left(\log\left|\frac{z-x_1}{z-x_2}\right|^2\bar{k}_{x_3x_4}+\log\left|\frac{z-x_3}{z-x_4}\right|^2\bar{k}_{x_1x_2}\right)\log(\bar{z}-\bar{x}_5)\right]\right.\\\left.-\frac{dz}{2\pi i}\left[\left(\log\left|\frac{z-x_1}{z-x_2}\right|^2k_{x_3x_4}+\log\left|\frac{z-x_3}{z-x_4}\right|^2k_{x_1x_2}\right)\log(z-x_5)\right]\right\},
\end{multline}
where we get an extra minus sign, since we changed the direction of the contour integrals to be counterclockwise. In the limit in which the width of the strip tends to zero, we obtain a line integral from $x_5-\infty$ to $x_5$. Evaluating the contour integrals around the $x_i$ is simple. For example, the integral around $x_1$ gives after setting $z=x_1+\epsilon e^{i\varphi}$ the finite contribution
\beq
\frac{1}{2 \pi i}\int_{0}^{2 \pi}(-i )d\varphi \log\left|\frac{x_{13}}{x_{14}}\right|^2\left(\log(\bar{x}_{15})+\log(x_{15})\right)=-\log\left|\frac{x_{13}}{x_{14}}\right|^2\log|x_{15}|^2,
\eeq
with the next term being proportional to $\epsilon\log(\epsilon)$. The expressions for $x_2$, $x_3$ and $x_4$ are similar, while the contour integral around $x_5$ gives no divergent or finite contributions. The integrand over the strip on the other hand gives a total derivative so that we get the expression
\beq
\int_{-\infty}^{-\epsilon}dy\, 2\text{Re}\left(\log\left|\frac{x_{51}+y}{x_{52}+y}\right|^2\left(\frac{1}{x_{53}+y}-\frac{1}{x_{54}+y}\right)+(12\leftrightarrow 34)\right)=\log\left|\frac{x_{15}}{x_{25}}\right|^2\log\left|\frac{x_{35}}{x_{45}}\right|^2,
\eeq
up to terms that vanish as $\epsilon$. Adding up everything,  we obtain the result
\begin{multline}
\label{eq:intlog2}
\calH^{x_1,x_2,x_3,x_4}_{x_5}=-\log\left|\frac{x_{13}}{x_{14}}\right|^2\log|x_{15}|^2+\log\left|\frac{x_{23}}{x_{24}}\right|^2\log|x_{25}|^2\\-\log\left|\frac{x_{13}}{x_{23}}\right|^2\log|x_{35}|^2+\log\left|\frac{x_{14}}{x_{24}}\right|^2\log|x_{45}|^2+\log\left|\frac{x_{15}}{x_{25}}\right|^2\log\left|\frac{x_{35}}{x_{45}}\right|^2+\calO(\epsilon\log(\epsilon)).
\end{multline}
Let us perform a check of the above formula. If we set $x_1=xL$ and $x_2=L$ and take $L$ to infinity, then eq.~\eqref{eq:intlog1} goes to zero independently of $x$, as long as it is not zero. Plugging the same limit in
the result~\eqref{eq:intlog2}, we get $\calH=-\log|x|^2\log|x_{35}|^2+\log|x|^2\log|x_{45}|^2+\log|x|^2\log\left|\frac{x_{35}}{x_{45}}\right|^2=0$ as required.

We now have all the tools require to compute the integral of the last term of eq.~\eqref{eq:3lintcc}. We need to evaluate the expression
\begin{multline}
\label{eq:calFpremier}
\calF\colonequals\int_{\calD_3}\frac{d^2z_1d^2z_2d^2z_3}{\pi^3}|h_3|^2\left[|h_2|^2\left(\frac{1}{z_1}-\frac{1}{z_{12}}
\right)\left( \frac{1}{\bar{z}_{13}}-\frac{1}{\bar{z}_1+1}
\right)+\right.\\\left.+h_2\bar{h}_1\left(\frac{1}{z_{12}}-\frac{1}{z_{13}}
\right)\left(\frac{1}{\bar{z}_{23}}-\frac{1}{\bar{z}_2+1}
\right)+\mathrm{c.c.}\right],
\end{multline}
where we have used the change  of variables $z_i\rightarrow u+(u-v)z_i$, which transforms the cut-off as  $\epsilon\rightarrow \frac{\epsilon}{|u-v|}$. This means that in eq.~\eqref{eq:calFpremier} as well as in the following formulas we have $h_i=\frac{1}{z_i}-\frac{1}{z_i+1}$. Integrating over $z_1$ using formula~\eqref{eq:basicintegral3}, we get
\beq
\calF= \int_{\calD_2}\frac{d^2z_2d^2z_3}{\pi^2}|h_3|^22\text{Re}\left[|h_2|^2\log\left|\frac{z_{23}}{(z_2+1)z_3}\right|^2+h_2\left(\frac{1}{\bar{z}_{23}}-\frac{1}{\bar{z}_2+1}\right)\log\left|\frac{(z_2+1)z_3}{z_2(z_3+1)}\right|^2\right].
\eeq
If we now integrate over $z_2$ by using eq.~\eqref{eq:basicintegral3} as well as eq.~\eqref{eq:intlog2}, we get
\begin{multline}
\calF=\int_{\calD_1}\frac{d^2z}{\pi}|h|^2\Big[\calH^{0,-1,0,-1}_z-\calH^{0,-1,0,-1}_{-1}+\calH^{0,-1,z,-1}_{-1}-\calH^{0,-1,z,-1}_0\\+4\log\epsilon^2\log|z|^2+2\log\left|\frac{z}{z+1}\right|^2\log\left|\frac{z+1}{z\epsilon}\right|^2\Big],
\end{multline}
where we have set $z\equiv z_3$. A priori, $\calH$ is singular whenever two $x_i$ are equal. One can compute the regularized result in two ways - either by performing a computation similar to the one that led to eq.~\eqref{eq:intlog2}, or by simply setting $x_{ij}=\epsilon$ when $x_i$ approaches $x_j$. Either way, we obtain
\begin{align}
\calH^{0,-1,0,-1}_z&=-2\log(\epsilon^2)\log\left(|z|^2|z+1|^2\right)+\left[\log\left|\frac{z}{z+1}\right|^2\right]^2,
\nonumber\\ \calH^{0,-1,0,-1}_{-1}&=-\left[\log(\epsilon^2)\right]^2,\nonumber\\
\calH^{0,-1,z,-1}_{-1}&=-\left[\log(\epsilon^2)\right]^2-\log|z+1|^2\log\left|\frac{z}{z+1}\right|^2,\nonumber\\
\calH^{0,-1,z,-1}_0&=-\log|z|^2\log\left|\frac{z}{z+1}\right|^2.
\end{align}
Putting everything together, we obtain the somewhat surprising result
\beq
\calF=0.
\eeq
Hence, up to and including 3-loops the structure constants do not contribute to the 2-point functions of affine primaries.

%%%%%%%%%%%%%%%%%%%%%%%%%%%%%%%%%%%%%%%%%%%%%%%%%%%%%%%%%%%%%%%%%%%%
%bibliography
%%%%%%%%%%%%%%%%%%%%%%%%%%%%%%%%%%%%%%%%%%%%%%%%%%%%%%%%%%%%%%%%%%%%

\end{document}